\documentclass{aa}

\usepackage{epsfig}
\usepackage{natbib}
\bibpunct{(}{)}{;}{a}{}{ }

\begin{document}

\title{X-ray Orbital Modulations in Intermediate Polars}
 
\author{ T.L. Parker \inst{1} \and A.J. Norton\inst{1} \and K. Mukai\inst{2} }

\offprints{T.L.Parker}

\institute{Department of Physics and Astronomy, The Open University, 
	Walton Hall, Milton Keynes MK7 6AA, U.K. \\
	\email{T.L.Parker@open.ac.uk, A.J.Norton@open.ac.uk,}
 \and
	NASA Goddard Space Flight Center, Laboratory for High Energy 
	Astrophysics, Code 662, Greenbelt, MD 20771, USA {\it and} Universities
	Space Research Association \\
	\email{mukai@milkyway.gsfc.nasa.gov}
}

\date{Received 2005 February 16;
      accepted 2005 March 19}

\abstract{ 
We present an analysis of 30 archival {\em ASCA} and {\em RXTE} X-ray 
observations of 16 intermediate polars to investigate the nature of their 
orbital modulation. We show that X-ray orbital modulation is widespread 
amongst these systems, but not ubiquitous as indicated by previous studies 
that included fewer objects. Only seven of the sixteen systems show a clearly 
statistically significant modulation depth whose amplitude decreases with 
increasing X-ray energy. Interpreting this as due to photoelectric absorption 
in material at the edge of an accretion disc would imply that such modulations 
are visible for all system inclination angles in excess of 60 degrees. However,
it is also apparent that the presence of an X-ray orbital modulation can
appear and disappear on a timescale of $\sim$years or months in an individual 
system. This may be evidence for the presence of a precessing, tilted
accretion disc, as inferred in some low mass X-ray binaries.

}

\maketitle

\section{Introduction}

Intermediate Polars (IPs) are semi-detached interacting asynchronous binaries 
in which a magnetic white dwarf (WD) accretes material from a Roche 
lobe-filling, usually main sequence dwarf, companion. The accretion flow from 
the secondary proceeds towards the WD either through an accretion disc, an 
accretion stream, or some combination of both (known as disc 
overflow accretion), until it reaches the magnetospheric radius. Here the 
material attaches to the magnetic field lines and follows them towards the 
WD magnetic poles where the infalling material takes the form of arc-shaped 
accretion curtains, standing above the WD surface. At some distance from this
surface, the accretion flow undergoes a strong shock, below which material 
settles onto the WD, releasing X-ray to optical emission. Since the magnetic 
axis is offset from the spin axis of the WD, this gives rise to the defining 
characteristic of the class, namely X-ray (and usually optical) emission 
pulsed at the WD spin period. X-ray pulsation at the lower orbital sideband 
frequency is sometimes seen and believed to be an intrinsic modulation arising 
from pole-switching in the case of stream-fed accretion. Over twenty confirmed 
IPs are now recognised, with a similar number of candidate systems having been 
proposed. Comprehensive reviews of various aspects of their behaviour are 
given by \citet{PAT94}, \citet{WAR95}, and \citet{NWS04}.

Orbitally modulated X-ray flux has been reported in several IPs and three 
possible origins for such X-ray orbital modulations have been suggested. In 
the absence of an accretion disc, the accretion onto the WD poles may depend 
on binary phase, and so give rise to an orbitally modulated X-ray flux. A 
second possibility is that the interaction of stream and disc could give rise 
to an additional emission component, whose visibility varies with orbital 
phase. Finally, if a disc is present with a non-axisymmetric structure at its 
outer edge, local absorption of the X-ray flux could occur as the structure 
crosses the line of sight to the WD. This would reduce the X-ray flux 
periodically, with an energy dependence characteristic of photoelectric 
absorption. Absorption at the edge of a disc, is the accepted explanation for 
X-ray orbital modulations seen in Low Mass X-ray Binaries (LMXBs) and is a 
likely candidate to operate in IPs too. 

\citet{HGM93} carried out the only previous systematic study of X-ray orbital 
modulations in IPs. They looked at the X-ray lightcurves of seven IPs observed 
by {\sl EXOSAT}, and found orbital modulations to be probably present in each 
of them (TV Col, TX Col, AO Psc, FO Aqr, BG CMi, V1223 Sgr and EX Hya), 
although the evidence in V1223 Sgr was marginal. \citet{HGM93} concluded that 
these orbital modulations are similar to those seen in dipping LMXBs and have 
a similar cause. The implication noted by Hellier et al. is that IPs are all 
seen at high inclinations and are all disc fed. For comparison, our 
observations of IPs using {\em ROSAT} and \textit{Ginga} confirmed the 
presence of X-ray orbital modulations in FO Aqr \citep{NWKLMc92}, BG CMi 
\citep{NMcLW92}, TX Col (Norton et al. 1997), and AO Psc \citep{TBNOW97}, and 
failed to find such modulation in V1223 Sgr \citep{TBNOW97}, YY Dra and V709 
Cas \citep{NBAH99}. In addition XY Ari and EX Hya are the only IPs to show 
X-ray eclipses \citep{HEL97, RMMW91} and so are known to be at high 
inclination angles. 

In order to extend the earlier studies, and investigate the preponderance of 
X-ray orbital modulations in the large number of IPs now known, we have 
re-analysed 30 X-ray observations of 16 IPs obtained with {\em RXTE} 
and {\em ASCA} that are available in the archives. We have excluded 
observations of IPs that span less than two orbital periods of the system in 
question and we have not included observations of EX Hya and XY Ari in our 
analysis. The X-ray orbital modulation of EX Hya is well studied but 
difficult to disentangle from the spin modulation with which it is in a 
3:2 resonance. The archival {\em RXTE} observations of XY Ari are all short 
and cover only a tiny phase range around the eclipse egress in each case, so 
the overall broad X-ray modulation (antiphased with the X-ray eclipse) is not 
sampled.

\section{Observations} 

The data presented here were obtained with the Advanced Satellite for 
Cosmology and Astrophysics ({\em ASCA}) and the Rossi X-ray Timing Explorer 
satellite ({\em RXTE}). Table 1 shows details of the {\em ASCA} observations 
and Table 2 lists the {\em RXTE} observations that we have used. References
in each table indicate papers where these observations have been previously
reported, although in most cases, the authors do not comment on any 
X-ray orbital modulation.  

{\em ASCA} performed imaging and spectroscopic observations simultaneously 
over an energy range of 0.5 -- 10~keV and carried four large-area X-ray 
telescopes. At the focus of two of the telescopes were Gas Imaging 
Spectrometers (GIS) with an energy resolution of $E/\Delta E$ $\sim$13 at 
6~keV and $\sim$7 at 1.5~keV, and a circular field of view of 50$^{'}$ 
diameter. Solid-state Imaging Spectrometers (SIS) with an energy resolution 
of $\sim$50 at 6~keV and $\sim$20 at 1.5~keV, and a square field of view of 
20$^{'}$ x 20$^{'}$, were at the focus of the other two. See \citet{TIH94} 
for a more extensive guide to {\em ASCA}. 

{\em RXTE} is a satellite designed for time domain studies of cosmic X-ray 
sources. The instrument used for the observations presented here was the 
Proportional Counter Array (PCA; see \citet{ZGJSSM93}), which consists of 
five Proportional Counter Units (PCUs) with a total net area of 
6500~cm$^{2}$. Each PCU is filled with xenon gas and split into three 
layers, plus a propane filled veto layer. The energy bandwidth of the 
instrument is 2--60~keV and it has a collimated field of view of 1$^{\circ}$ 
(FWHM). The energy resolution of the PCA is $\sim$18\% at 6~keV 
\citep[see][ for more details]{BRS93}. During the observations reported 
here, the PCA was mostly operated with all five PCUs functioning, however for 
some observations fewer were in use, but this is accounted for within the 
data reduction and the lightcurves are shown as counts per PCU against time.

\section{Data reduction}

All the data used were extracted from the HEASARC archive and reduced using 
the standard routines of the {\sc ftools} package v5.2.

For the {\em RXTE} data reduction, we took the standard data products from 
the PCA and extracted the source lightcurves for each dataset using the tool 
{\sc saextract} for the appropriate good time intervals. The background 
data files were generated with the tool {\sc runpcabackest} using background 
models appropriate for the source. We filtered out inappropriate or 
contaminated data by applying the criterion that the target elevation $>$5, 
pointing offset $<$0.02, time since passage through South Atlantic Anomaly 
$>$25~min, and used a standard threshold for electron contamination. 
Background subtracted lightcurves were then constructed in 16~s time bins 
in the energy bands 2--4, 4--6, 6--10 and 10--20~keV.

ASCA data reduction was carried out in a similar way, with data from all four 
co-aligned instruments. The reduction was carried out using standard 
screening criteria (using {\sc ascascreen}), filtering data for Earth 
elevation angle, stable pointing, the South Atlantic Anomaly passages, and 
cutoff rigidity. Hot and flickering pixels were removed from the SIS events 
and SIS data were additionally screened for proximity to the bright Earth. 
Only well-calibrated events, with grades 0, 2, 3 or 4 were kept. GIS data were 
additionally screened to remove the background ring, and calibration source. 
Light curves were extracted in 16~s bins from each instrument in low 
(0.7--2~keV) and high (2--10~keV) energy bands, both from the source and from 
nearby source-free background region using {\sc xselect}. Background 
subtraction was carried out and data from all four instruments were then 
combined accounting for different good times.  

Lightcurves from both {\em RXTE} and {\em ASCA} data were then folded into 
16 orbital phase bins according to the best available orbital ephemerides for 
each system (see Table 3). In each case the epoch of zero phase was adjusted 
to coincide with the presumed inferior conjunction of the secondary star. The 
published ephemerides of these systems are defined in various ways. If the 
ephemeris is based on optical photometric data, we assume that minimum light 
corresponds to the secondary star at inferior conjunction, or that maximum 
light indicates the WD at inferior conjunction. Similarly, if the ephemeris is 
based on spectroscopic data, we assume that red-to-blue crossing of the 
emission lines corresponds to the WD at superior conjunction, or that 
blue-to-red crossing of the absorption lines indicates the secondary star at 
inferior conjunction. Where no ephemeris is available, the phasing is 
arbitrary and zero phase is simply set to the start of the observation in 
each case. 

Power spectra of each lightcurve were constructed using the 1-dimensional 
{\sc clean} algorithm of H.J. Lehto, which utilizes a variable gain. This is 
particularly suited to revealing the signals in unevenly sampled, 
multiply-periodic data and is quite effective at removing signals at the 
frequency $1.75 \times 10^{-4}$~Hz due to the sampling pattern imposed by the 
orbital period of both the {\em ASCA} and {\em RXTE} satellites (about 95 
minutes in each case). Noise levels in {\sc clean}ed power spectra are 
notoriously difficult to calculate \citep{NBT96}. However, here we are in the 
fortunate position of knowing the expected system frequencies in advance, 
namely the various harmonics and sidebands of the spin and orbital frequencies 
of each system. Any power at a frequency other than those `expected' can be 
considered spurious, and due to noise in the lightcurve. In order to estimate 
the noise level in a given power spectrum we therefore follow the prescription
of \citet{NBT96}. This involves determining the power level above which there 
are $n$ discrete noise peaks in the power spectrum, where we define $n$ in
this case as 1\% of the total number of independent frequency samples in the 
spectrum (typically several thousand, depending on the length and sampling 
of that data set). This power level therefore represents that at which there 
is a 1\% chance that any individual frequency bin will contain a noise peak. 
Hence, if the power at any of the expected system frequencies is greater than 
this noise level, there is a 99\% chance that the signal at that 
frequency is {\em not} due to noise. Such peaks are therefore considered
as `real'. Conversely, any peak that corresponds to one of the system 
frequencies, but which has a power below the noise level, is considered as 
spurious. This simple procedure allows us to make a direct comparison between 
one data set and another. Power spectra for a sample set of objects are 
shown in Figures 1 and 2 (the full set are available in the on-line version),
with the noise level indicated by a dashed horizontal line in each case.

Finally, in order to quantify the degree of X-ray orbital modulation in each 
system, we found the best-fit sinusoid to each folded lightcurve, allowing 
the mean level, amplitude and phase as free parameters. We recognise that many
of the orbital modulations present may not be well fit by a sinusoid, but
this is the simplest way of characterising the depth in each system in a 
systematic manner. The folded lightcurves for a sample set of objects
are shown in Figures 1 and 2 (the full set are available in the on-line
version) with the best-fit sinusoids over-plotted in each case. Modulation 
depths were calculated as the peak-to-peak amplitude of the sinusoid divided 
by the maximum level, and are shown in Tables 4 and 5, along with the fitted 
phase of minimum flux and the reduced chi-squared values for the sinusoid fit
in each case. 1$\sigma$ uncertainties on each parameter in the 
fit were calculated by rescaling the error bars on individual flux 
measurements so as to give a reduced chi-squared of unity for the best fit, 
and then finding the parameter range corresponding to an increase in reduced 
chi-squared of one. Figures 3a and 3b illustrate the data from Tables 4 
and 5 respectively. An X-ray orbital modulation is considered to be confirmed 
only if the depth is non-zero above the $3\sigma$ level in all energy bands 
of a given observation, and if the value for the phase of minimum flux is 
consistent between the various energy bands.

\section{Results}

Below we discuss the detailed results for each of the systems, listed in 
order of increasing orbital period.

\subsection{V1025 Cen}

V1025 Cen was discovered by \citet{BCRW98} and found to have a spin period 
of 2147s and a suggested orbital period of 1.41hr \citep{HBB98}. This object
is little studied, with no published ephemeris and no 
determination of the binary inclination. From the {\em ASCA} data (Fig. 1a) 
no significant orbital modulation is seen in the phase folded lightcurves 
though there is some variability. The power spectra show signals only at
the WD spin frequency and its harmonics. From the sinusoid fits to the 
orbital folded data we see that the modulation depths are consistent with 
zero (Table 4) in both energy bands. The {\em RXTE} data (Fig. 2a), like 
the {\em ASCA} data, show a variability in the folded lightcurves but the 
power spectra show nothing to indicate a clear orbital modulation. The 
sinusoid fits however, show evidence for a modulation depth that is just 
significant at the $\sim 3 \sigma$ level at all energies (Table 5) and this 
is reinforced by the fitted phase of minimum flux being consistent with the 
same phase across the four energy bands.

\subsection{BG CMi}

BG CMi is a well-known IP with an orbital period of $\sim$ 3.25hr and a spin 
period of 913s. The published orbital ephemeris of \citet{HEL97} is referenced 
to a spectroscopic eclipse believed to coincide with inferior conjunction of 
the secondary. BG CMi has a moderate inclination angle, 
$\sim 55^{\circ}-75^{\circ}$ inferred by \citet{deMMB-BVRMAG95} from the 
orbital modulation. The {\em ASCA} observation of BG CMi comprised two 
separate visits, separated by only a few days, in order to fully sample the
orbital phase of the system whose period is close to twice the orbital period
of the {\em ASCA} satellite. The resulting phase folded lightcurve (Fig. 1b) 
shows a strong dip at phase 0.7 in both energy bands and the corresponding 
power spectra reveal the expected strong signals at the orbital frequency.
Sinusoid fits to the modulation yield large modulation 
depths which decrease with increasing energy. The {\em RXTE} 
phase folded lightcurves (Fig. 2b) also show a modulation dip at 
phase 0.7 in all energy bands, with corresponding power spectra peaks 
at both the orbital and twice the orbital frequency. From the sinusoid 
fitting, the modulation depths show a roughly constant amplitude at all 
energies.

\subsection{V1223 Sgr}

V1223 Sgr is amongst the brightest IPs in the 1--20~keV band, with an orbital 
period of 3.36hr \citep{JS87}, and a spin period of 745.63s \citep{ORMB85}. 
The published orbital ephemeris of \citet{JS87} is referenced to photometric 
maximum. \citet{WGGHB85}, suggest that the inclination of this system is 
between 16$^{\circ}$ and 40$^{\circ}$. For this system a clear modulation 
dip is seen in the phase folded lightcurve of the {\em ASCA} data (Fig. 1c) 
at phase 0.65 and clearly seen in the power spectra with peaks at 
the orbital frequency and also at twice the orbital frequency. 
The sinusoid fits confirm a decreasing modulation depth with increasing 
energy. The {\em RXTE} data give a similar outcome (Fig. 2c). Although only 
a slight variation is seen in the folded lightcurves, a shallow modulation 
dip is apparent at the same phase as seen by {\em ASCA}. A correspondingly 
small peak at the orbital frequency is seen in the power spectra, though a 
lot of noise is present. The sinusoid fits yield small modulation depths 
that are roughly constant in all energy bands, or possibly even increase 
slightly with energy, and are significant across all four energy bands.

\subsection{V2400 Oph}

The first discless IP \citep{BSMOCSPH95, BHMPSS97}, has a spin period of 
927s and an orbital period of 3.42hr, although no ephemeris has been published 
for V2400 Oph. It is believed to be at a low inclination of 10$^{\circ}$ 
\citep{HB02}. Observed with {\em ASCA} it shows little in the way of an 
orbital modulation in the phase folded lightcurve (Fig. 1d). The power spectra 
and sinusoid fit reveal a similar result, with only a possible peak close to 
the orbital frequency and a modulation depth that is consistent with 
zero. A modulation is more apparent in the {\em RXTE} observation (Fig. 2d). 
There is a clear variability in the phase folded lightcurves, with a shallow
dip at (arbitrary) phase 0.7. Although no clear peaks are seen in the 
power spectra at the orbital frequency, the modulation depths from sinusoid 
fitting are small but just significant at the $3 \sigma$ level and roughly 
constant across the energy bands, or possibly even increasing with energy as 
seen in V1223~Sgr. 

\subsection{AO Psc}

AO Psc has a spin period of 805s and an orbital period at 3.59hr \citep{WM81,
PP81}. The orbital ephemeris of \citet{KS88} is referenced to 
photometric maximum. AO Psc has features very similar to that of V1223 Sgr 
from photometric observations and the star's optical spectrum \citep{PP81}. 
This similarity is also evident in the folded lightcurves and power spectra 
of the {\em ASCA} observation (Fig. 1e). A strong modulation is clearly seen 
with a minimum close to phase 0.1, and clear peaks in the power spectra 
are seen at the orbital frequency. The fitted modulation depths
decrease with increasing energy. With the {\em RXTE} observation, the 
situation is quite different (Fig. 2e). There is no strong modulation in the 
phase folded lightcurve and hardly any variability, and no orbital peaks in 
the power spectra, which are essentially delta functions at the WD spin 
frequency. The fitted modulation depths are not significant above the 
$3 \sigma$ level. This system shows clear evidence that the orbital 
modulation in IPs can come and go on a timescale of a few years.

\subsection{YY Dra}

YY Dra is one of a small number of IPs that show outburst behaviour. 
\citet{PSPBWC92} found the spin period to be 529s, while \citet{MSG91} found 
the orbital period at 3.96hr and derived the inclination to be 
$42^{\circ}\pm 5^{\circ}$. The orbital ephemeris of \citet{HPTHS97} is 
referenced to the inferior conjunction of the mass donor. YY Dra's phase 
folded lightcurve from the {\em ASCA} data does show some variability, and 
this is reflected in the power spectra with a peak at twice the orbital 
frequency (Fig. 1f). However, the sinusoid fits reveal no significant 
modulation. The {\em RXTE} data too indicate only slight variability (Fig. 2f).
This is confirmed in the power spectra where no modulations are seen at 
either the orbital or twice the orbital frequency. The sinusoid fits are 
consistent with zero modulation depth in all bands. 

\subsection{LS Peg}

LS Peg has an orbital period of 4.19hr \citep{MRC99, TTP99}, and 
a claimed spin period of 1776s \citep{RCMHS01} deduced from polarimetry. 
Recent analysis of the {\em ASCA} data on this object by \citet{BWO2005}
has detected a modulation at 1854s which is probably a more accurate 
determination of the spin period, and confirms this object as a genuine IP.
The orbital ephemeris of Taylor et al. (1999) is based on the blue-to-red 
crossing of the emission 
lines, which we interpret as inferior conjunction of the white dwarf. Only 
observed with {\em ASCA}, this system shows a strong variability in its 
orbital phase folded lightcurve (Fig. 1g). However, the modulation depths 
are not significant above the $3 \sigma$ level. There are peaks close to the 
orbital frequency and twice the orbital frequency, particularly in the lower 
energy band power spectrum. Although both power spectra are noisy, signals
related to the recently-identified spin period are clearly seen in both
bands.

\subsection{V405 Aur}

\citet{HTMSPSP94} found the orbital period of this system to be 4.15hr, and 
\citet{SKI96} discovered the white dwarf spin period of 545.46s. The 
published orbital ephemeris, cited in \citet{HH99}, is defined at inferior 
conjunction of the emission-line source (blue-to-red  crossing) and a 
correction of 0.1 cycles anticlockwise is claimed to give the inferior 
conjunction of the secondary. The first of the two observations taken with 
{\em ASCA} shows a relatively unvarying phase folded lightcurve (Fig. 1h)
with no modulation significant above the $3 \sigma$ level, although
the apparent phase of minimum flux is consistent at 0.7 between the two energy
bands. The power spectra show no evidence for an orbital signal either. 
However, the second {\em ASCA} observation (Fig. 1i) shows a prominent dip 
in the folded lightcurves and the power spectra indicate significant power at 
both the orbital frequency and twice the orbital frequency. The sinusoid 
fits to these folded lightcurves struggle to fit the sharp dip in the low 
energy lightcurve near to phase 0.4 and instead find the sinusoidal minimum 
close to phase 0.55, in both energy bands. However, only the modulation 
depth in the low energy band is significant above the $3 \sigma$ level.
The {\em RXTE} data show 
something a little different (Fig. 2g). The folded lightcurves have a very 
slight dip in each energy band around phase 0.7--0.8, with a significant
modulation, whose depth is roughly constant with energy. The power spectra 
indicate a small peak close to the orbital frequency in each case.

\subsection{V2306 Cyg}

The orbital period of V2306 Cyg is 4.35hr \citep{ZTE02} and its 
spin period is 1466.66s \citep{NQKLMN02}. The published orbital ephemeris is 
referenced to inferior conjunction of the secondary from both photometry and 
spectroscopy \citep{ZTE02}. Another system to be observed only by 
{\em ASCA}, V2306 Cyg is the recently adopted name of 1WGA~J1958.2+3232. The 
phase folded lightcurve of this system is an interesting one with a rather 
variable flux and an apparent modulation dip seen at phase 0.5 (Fig. 1j). If 
we compare with the power spectra, in the low energy band a marginal peak at 
twice the orbital frequency is seen with also a small peak associated with the 
orbital frequency. If we move to the higher energy band, the orbital frequency 
peak becomes more visible. The sinusoid fits are consistent with a decreasing 
modulation depth with increasing energy but the modulation is not seen above
$3\sigma$ significance in either band.

\subsection{FO Aqr}

\citet{PKRSVJBOK98} found the orbital and spin periods to be 4.85hr and 1254s 
respectively with the ephemeris referred to the optical photometric minimum. 
This system shows a very prominent dip in the phase folded lightcurve from 
the {\em ASCA} data (Fig. 1k) in both energy bands. The power 
spectra show that the orbital frequency and its harmonic are detected strongly 
and the sinusoid fits confirm the decreasing modulation depth with increasing 
energy. The phase folded {\em RXTE} lightcurves also show a clear and 
prominent modulation in all four energy bands and the power spectra confirm 
this. The modulation depths determined from sinusoid fits also confirm 
what is shown in the figures, with a steadily decreasing modulation depth with 
increasing energy. The fitted phases of minimum flux suffer from the 
modulation profile being distinctly non-sinusoidal: in both observations, 
there is a deep minimum followed by a shallower minimum. In the {\em ASCA} 
data these minima occur at phases 0.9 and 0.3 respectively, whereas in the 
{\em RXTE} observation they have apparently shifted to phase 0.75 and 0.25.
This apparent shift may be due to the shape of the modulation profile 
changing with energy, or alternatively it may be evidence for a changing 
accretion structure on a four year timescale as the accumulated phase error 
from the ephemeris is less than 0.02 in phase.

\subsection{PQ Gem}

The ephemeris for PQ Gem \citep{HEL97} is defined relative to red-to-blue 
crossing of the emission-lines. The system has an orbital period of 5.19hr 
and a spin period of 833.41s \citep{HEL97}. In the first observation of PQ Gem 
with {\em ASCA} (Fig. 1l), the power spectra show a peak associated with the 
orbital frequency in the low energy band, but nothing is seen in the second
observation (Fig. 1m). The folded lightcurves of both observations are 
unconvincing and although the modulation depths from the first observation
are just significant at the $3 \sigma$ level, those from the second observation
are not. Moving to the {\em RXTE} data (Fig. 2i), there is again little 
evidence of orbital modulation in either the folded lighturves or power 
spectra, and the fitted modulation depths are once again consistent with zero 
at the $3 \sigma$ level.

\subsection{V709 Cas}

Both the ephemeris and the orbital period for this system were determined 
by \citet{BMdeMMM01}, the ephemeris being with respect to the blue-to-red 
crossing time of emission lines and the orbital period is 5.34hr. The spin 
period of the white dwarf is known to be 312.8s \citep{HM95, NBAH99}. V709 
Cas has been observed by {\em RXTE} only (Fig. 2j). It shows no evidence of 
an orbital modulation in either its folded lightcurves or power spectra, and 
the fitted modulation depths are all consistent with zero.

\subsection{TV Col}

TV Col has an orbital period of 5.486hr \citep{HCCTC81} and a spin period of 
1911s \citep{SBvWWK85} as well as a second photometric (superhump) period
and a longer disc precession period. The orbital ephemeris \citep{HEL93} is 
referenced to a shallow photometric eclipse. TV Col shows very interesting 
features in both observations. The {\em ASCA} folded lightcurves (Fig. 1n) 
give a very clear and strong modulation in both energy bands with a 
minimum close to phase 0.0. The {\em RXTE} data reveal a similar effect 
(Fig. 2k) and although the modulation from the {\em RXTE} data is not as 
pronounced as that seen from {\em ASCA}, it is clearly present at low 
energies with a minimum around phase 0.9. The fitted modulation
depths in both observations show the classic trend of decreasing depth with 
increasing energy. The power spectra of the {\em ASCA} data show strong peaks 
associated with the orbital frequency as do the {\em RXTE} power spectra, 
particularly at lower energies. As with FO Aqr, the apparent discrepancy 
between the phases of minimum flux in the two observations may be due to the
changing shape of the modulation profile with energy and the inadequacy of a 
sinusoidal fit.

\subsection{TX Col}

TX Col is a supposedly a low inclination system $<30^{\circ}$ \citep{MBBT91}  
with an orbital period of $\sim$5.72hr and a spin period of 1911s \citep{BT89},
but no ephemeris has been published for it. TX Col reveals a strong modulation 
in the phase folded lightcurves of both satellites' observations and in both 
the depth decreases with increasing energy. In the {\em ASCA} data (Fig. 1o) 
the dip occurs at (arbitrary) phase 0.2 whilst in {\em RXTE} it occurs 
at (arbitrary) phase 0.9. The power spectra of the two observations
confirm what is seen in the folded lightcurves, each showing a strong peak at 
the orbital frequency.

\subsection{AE Aqr}

The magnetic propellor system AE Aqr is a very unusual IP with a long orbital 
period of 9.88hr \citep{WHG93} and a short spin period of 33s \citep{deJMOR94}.
The spectroscopic orbital ephemeris is referenced to superior conjunction of 
the white dwarf. Only observed by {\em ASCA}, the X-ray lightcurve exhibited
strong flares during the observation. However, even after removing these 
sections, the phase folded lightcurve shows a large variation. The sinusoid 
fits imply a relatively large modulation depth in both bands with a consistent 
phase of minimum flux, but  both are significant at only just above the 
$3 \sigma$ level. The power spectra are very noisy at 
low frequencies although there are indications of peaks coincident with both 
twice the orbital frequency and the orbital frequency itself in each case.

\subsection{V1062 Tau}

This little studied system has no published ephemeris and no inclination 
angle is known. The spin period and orbital period are 3726s \citep{HBM02} 
and 9.95hr \citep{RBBBSSST94} respectively. The {\em ASCA} and {\em RXTE} 
data both show strong variability (Fig. 1q and 2n) and although sinusoid fits 
reveal modulation depths that either decrease with increasing energy or 
remain constant, these modulations are not significant above the $3 
\sigma$ level in any of the energy bands. No clear orbital modulation 
frequency is seen in any of the power spectra.

\section{Summary of results}

Of the sixteen IPs studied, seven systems show evidence for an X-ray 
orbital modulation whose amplitude decreases with increasing X-ray energy, 
in at least one observation. These are: BG CMi, FO Aqr, TV Col and TX Col in 
their observations with both {\em ASCA} and {\em RXTE}, and  
V1223 Sgr, AO Psc and V405 Aur (second observation), each in only their 
{\em ASCA} observations but not their {\em RXTE} observations. 
Interestingly, six of these plus EX Hya comprise the seven objects 
studied by \citet{HGM93} in the original {\em EXOSAT} investigation of IP 
X-ray orbital modulation which found such modulation to be ubiquitous. This 
result also confirms the findings of the various {\em ROSAT} and {\em Ginga} 
studies of individual objects over the last decade.

A further six systems show X-ray orbital modulation that is just significant 
at the $3 \sigma$ level or whose depth is apparently constant across the 
energy bands in one observation. These are the {\em RXTE} observations of 
V1025 Cen, V1223 Sgr, V2400 Oph and V405 Aur, and the {\em ASCA} observations
of V2306 Cyg and AE Aqr. Given the noisy nature of some 
of these data and the large uncertainties on the measured modulation depths, 
many of these are also consistent with a slight decrease in modulation depth 
with increasing energy.

Finally, nine systems are consistent with zero X-ray orbital modulation, 
across the energy range, in at least one observation. These are: V1025 Cen, 
V2400 Oph, LS Peg and V405 Aur (first observation) in their {\em ASCA} 
observations; YY Dra, PQ Gem and V1062 Tau in both their {\em ASCA} and 
{\em RXTE} observations; and AO Psc and V709 Cas in their {\em RXTE} 
observations.
 
There is clearly a variety of orbital modulation depths seen in these systems.
In the lowest energy range (0.7 -- 2 keV, {\em ASCA}), the depths measured
at greater than $3 \sigma$ significance vary from 100\% (FO Aqr) to 28\% 
(V1223 Sgr). Similarly, in the highest energy range (10 -- 20 keV, 
{\em RXTE}), the largest modulation depth significantly detected is 34\% 
(BG CMi), whilst the smallest is 12\% (V1223 Sgr). The uncertainty in the 
modulation depth is generally of order a few percent, although sometimes as 
small as 1\%. Hence, the smallest modulation depth that can reliably be 
detected is typically around 10\%, although in a few cases we would be 
sensitive to detecting modulation depths of less than 5\%. However, in no 
instance here do we see a modulation depth as small as this detected with 
much more than $3 \sigma$ significance.

\section{Discussion}

From these results we conclude that X-ray orbital modulation in IPs is indeed 
widespread, but not ubiquitous. The dominant behaviour is that of decreasing 
modulation depth with increasing X-ray energy, suggesting photoelectric 
absorption as the cause. The most likely site for this additional absorption 
is in material thrown up at the edge of the accretion disc, due to impact by 
the accretion stream. As noted earlier, this is similar to the situation 
evisaged to cause X-ray dips in low mass X-ray binaries. The effect has also 
been realised in three-dimensional SPH simulations of cataclysmic variables 
by \citet{KSH01}, who show that dips will be caused around orbital phase 
0.7 if the system inclination angle is at least $65^{\circ}$.
 
We note that an additional column density of 
$N_{\rm H} = 10^{22}$~cm$^{-2}$ would produce modulation depths
of $\sim 80\%$ at 1keV, $\sim 10\%$ at 3keV, $\sim 5\%$ at 5keV and have 
negligable effect  above 8keV. Increasing the additional column density at 
X-ray minimum to $N_{\rm H} = 10^{23}$~cm$^{-2}$ would increase these 
modulation depths to $\sim 100\%$, $\sim 60\%$ and $\sim 20\%$ respectively, 
with significant modulation out to above 10keV \citep{NW89}. A simple partial 
covering absorber can reduce the modulation depth at low energies 
significantly, whilst still maintaining the same effect above $\sim 4$keV. 
The values in Tables 4 and 5 for those systems exhibiting a decreasing 
modulation depth with increasing energy suggest that additional column 
densities of this order may be present in many of these IPs, at least some 
of the time, at X-ray orbital minimum. A roughly constant modulation 
depth across a wide energy range, as seen in some systems, would imply 
that some fraction of the X-ray emission is completely blocked for part
of the orbit, akin to an occultation.

In total, 11 systems out of the 16 observed here exhibit a possible X-ray 
orbital modulation during at least one of the observations we have analysed. 
If we add in EX Hya and XY Ari which are known to have an X-ray eclipse, then 
the proportion of systems showing X-ray orbital modulation is 72\%. Assuming 
the systems to be randomly distributed in inclination angle, and that X-ray 
modulation is preferentially seen at higher angles, implies that an X-ray 
orbital modulation is visible for all inclination angles greater than 
around $44^{\circ}$. If instead we only recognise the X-ray orbital modulation 
seen above a $3 \sigma$ level throughout the energy range observed, namely 
that in BG CMi, V1223 Sgr, AO Psc, V405 Aur, FO Aqr, TV Col and TX Col 
(plus the two eclipsing systems), then the proportion
of systems falls to 50\% and the limiting inclination angle at which an 
orbital modulation is seen is increased to $60^{\circ}$, which is in  
good agreement with the predictions from the SPH simulations of 
\citet{KSH01}.

The orbital phase at which minimum X-ray flux occurs in different systems
is seen to be predominantly in range $\sim 0.7 - 1.0$. As noted above, stream 
impact with the outer edge of the disc is expected to occur around orbital 
phase 0.7, and this is indeed seen in BG CMi and V1223 Sgr (both observations 
of each), and FO Aqr ({\em RXTE} observation). Of the IPs with a well 
established orbital modulation and a previously published ephemeris, AO Psc
stands out in having a minimum after phase zero instead of before. However,
the accumulated phase error by the time of the {\em ASCA} observation is 
0.1 (see Table 3), so this may not be significant. In some of the systems we 
have studied, we see evidence that may be interpreted as a variation in the 
orbital phase at which minimum X-ray flux occurs from one observation to the
next, notably in FO Aqr and TV Col.
This may indicate that the absorbing material is not fixed in its location
at the edge of the disc. Alternatively, both the lack of dips centred near
phase 0.7 and the apparent variability in phase of X-ray minimum may 
simply reflect greater inaccuracies in the published orbital ephemerides
than claimed, or uncertainties in assigning a particular system geometry
to a given phase determined from a spectroscopic ephemeris.

We also note that the presence of X-ray orbital modulation in a given IP may 
come and go on a timescale of years or months. This is particularly apparent 
in two systems: for AO Psc a modulation was seen by {\em EXOSAT} (1983 and 
1985), {\em Ginga} (1990), {\em ROSAT} (1994) and {\em ASCA} (1994), but not 
by {\em RXTE} (1997); whereas for V1223 Sgr, a modulation was possibly seen by 
{\em EXOSAT} (1983 and 1984), not seen by {\em Ginga} (1991) and {\em ROSAT}
(1994), then seen by {\em ASCA} (1994) and {\em RXTE} (1998). We also note
the significant differences displayed by the two {\em ASCA} observations
of V405 Aur taken only two-and-a-half years apart.
This indicates that the visibility or size of whatever structure is 
responsible for producing the photoelectric absorption also varies 
considerably on this timescale. A similar effect is seen in some LMXBs
\citep{SMWG88, Schmidtke88} and is ascribed to the presence of a precessing, 
tilted accretion disc. Such a disc has been suggested as the cause of negative 
superhumps in cataclysmic variables \citep{ PATT99} although simulations vary 
in their success at reproducing the phenomenon \citep{Larwood98, 
MA98, WMS2000, MCWK02}. Nonetheless, if a tilted disc 
is present in an IP, a raised bulge at its outer edge caused by stream impact 
could move into and out of the line of sight to the X-ray source as its 
extent above the orbital plane varies with disc precession phase. In 
this way, a varying presence or depth of X-ray orbital modulation would 
naturally arise. The only IP with a confirmed tilted precessing disc is 
TV Col which has a 4~d disc precession period \citep{BOW88}, but it is 
conceivable that such a phenomenon may also be present in other systems.
It would be interesting to monitor the X-ray orbital modulation from 
TV Col throughout its 4-day disc precession period in order to test this
hypothesis.

Finally, we note that the apparently constant modulation depth as a function 
of energy seen in some IPs we have studied, at some epochs (i.e. {\em RXTE} 
observations of V1223 Sgr, V2400 Oph and V405 Aur), might indicate 
that an additional emission component is responsible for some or all of the 
variation in flux as a function of orbital phase. Alternatively, or 
additionally, it might imply that the absorbing region is highly structured or 
patchy and so the amount of absorption does not simply decrease with 
increasing energy. However, we see no constant or increasing modulation depth 
with increasing X-ray energy at a level much greater than $3\sigma$ 
significance so do not place too much weight on this interpretation.

\begin{acknowledgements} 
This research has made use of data obtained from the High Energy Astrophysics 
Science Archive Research Center (HEASARC), provided by NASA's Goddard Space 
Flight Center.  We thank the anonymous referee for several helpful suggestions
which have improved the paper.
\end{acknowledgements}

\bibliographystyle{aa}
\bibliography{2887}

\newpage

\clearpage

\begin{table}
\caption{ASCA observations}
\begin{tabular}{lccccc}
\hline
Target name  & Observation start time  & Duration / s  & GIS exposure / s  & SIS exposure / s & Ref.  \\
\hline
  V1025 Cen  &    1997-01-14 08:46:26&  131048&  60832&  48320   &--\\
  BG CMi     &    1996-04-14 18:17:33&  198956&  85440&  72352   &--\\
  V1223 Sgr  &    1994-04-24 04:15:47&  153904&  59168&  55552   &[1]\\
  V2400 Oph  &    1996-03-18 20:26:45&  206031&  83840&  76320   &--\\
  AO Psc     &    1994-06-22 06:52:43&  196679&  82336&  75808   &[2]\\
  YY Dra     &    1997-05-06 09:16:20&   80064&  32896&  24448   &[3]\\
  LS Peg     &    1998-11-23 11:21:58&   79728&  29936&  28176   &[4]\\
  V405 Aur (1) &  1996-10-05 13:36:29&  165263&  82368&  71323   &--\\
  V405 Aur (2) &  1999-03-21 09:24:39&   91570&  44064&  38272   &--\\
  V2036 Cyg  &    1998-05-15 03:18:51&   56816&  28352&  25248   &[5]\\
  FO Aqr     &    1993-05-20 22:12:34&   96176&  37952&  35552   &[6],[7]\\
  PQ Gem (1) &    1994-11-04 21:44:28&  189439&  79872&  71200   &[8],[9]\\
  PQ Gem (2) &    1999-10-19 16:23:56&   98800&  42912&  39296   &--\\
  TV Col     &    1995-02-28 06:03:42&  102998&  39936&  36448   &[10]\\
  TX Col     &    1994-10-03 07:04:37&   96992&  41888&  35680   &[11]\\
  AE Aqr     &    1995-10-13 23:40:13&   85842&  43968&  39264   &[12],[13]\\
  V1062 Tau  &    1998-02-16 21:33:55&  137264&  61248&  55136   &[14]\\
\hline
\end{tabular}
References: [1] \cite{BOH00}, [2] \cite{HMIF96}, [3] \cite{YB02}, [4] \cite{SNLF01}, [5] \cite{NQKLMN02}, [6] \cite{BMNOH98}, [7] \cite{MIO94}, [8] \cite{JRCB02}, [9] \cite{M97}, [10] \cite{RSSB04}, [11] \cite{NHBWOT97}, [12] \cite{E99}, [13] \cite{CDA99}, [14] \cite{HBM02}
\end{table}

\clearpage

\begin{table}
\caption{RXTE observations}
\begin{tabular}{lccc}
\hline 
  Target name  & Observation start time  &   Total exposure time / s & Ref.\\
\hline
  V1025 Cen &     1997-01-14 17:03:58.2&   89437 & -- \\
  BG CMi    &     1997-01-07 11:40:56.9&   98894 & -- \\
  V1223 Sgr &     1998-11-30 22:28:07.7&  136152 & [1]\\
  V2400 Oph &     1996-03-12 21:17:04.7&   89649 & -- \\
  AO Psc    &     1997-09-06 01:09:00.3&   94351 & [2]\\
  YY Dra    &     1996-03-13 08:57:09.8&   29243 & [3]\\
  V405 Aur  &     1996-04-26 00:50:55.1&   57899 & --\\
  FO Aqr    &     1997-05-14 19:01:23.2&  180380 & --\\
  PQ Gem    &     1997-01-27 03:54:38.0&  137874 &[2],[4]\\
  V709 Cas  &     1997-03-28 00:10:54.0&   59485 & [5]\\
  TV Col    &     1996-08-10 14:57:53.9&   83731 & [6]\\
  TX Col    &     1997-03-25 06:09:29.4&   68555 & --\\ 
  V1062 Tau &     1998-02-16 06:29:55.4&  146526 & [7]\\
\hline
\end{tabular}
\\
References: [1] \cite{RLSSZ04}, [2] \cite{S99}, 
[3] \cite{SNEMHUKPKSC02}, [4] \cite{JRCB02}, [5] \cite{deMMMBBCGHM01}, 
[6] \cite{RSSB04}, [7] \cite{HBM02} 
\end{table}

\clearpage

\begin{table}
\caption{Orbital ephemerides} 
\begin{tabular}{l r c c l }
\hline
Source & Orbital Ephemeris/HJD$^{\dag}$ & Ref. & Zero Phase/HJD                 & Accumulated phase error \\
       &                       &      & (inf. conj. of $2^{\rm ndary}$)& at epoch of X-ray observations \\
\hline 
\ V1025 Cen$^{*}$  & 0.05876                         &  [1]   & --          & --\\
\ BG CMi           & 2445020.396(2)+0.13474854(10)E  &  [2]   & 2445020.396 & {\em ASCA}(1): 0.03 ; {\em ASCA}(2): 0.03 ; {\em RXTE}: 0.03  \\
\ V1223 Sgr        & 2444749.986(3)+0.1402440(4)E    &  [3]   & 2444750.056 & {\em ASCA}: 0.10 ; {\em RXTE}: 0.13 \\
\ V2400 Oph$^{*}$  & 0.1429167                       &  [4]   & --          & -- \\
\ AO Psc           & 2444864.1428(28)+0.1496252(5)E  &  [5]   & 2444864.217 & {\em ASCA}: 0.10 ; {\em RXTE}: 0.13  \\
\ YY Dra           & 2446863.4376(5)+0.16537398(17)E &  [6]   & 2446863.4376& {\em ASCA}: 0.02 ; {\em RXTE}: 0.02 \\
\ LS Peg           & 2450358.8964(9)+0.174774(3)E    &  [7]   & 2450358.983 & {\em ASCA}: 0.08 \\
\ V405 Aur         & 2449474.6446(15)+0.172624(1)E   &  [8]   & 2449474.744 & {\em ASCA}(1): 0.3 ; {\em ASCA}(2): 0.6 ; {\em RXTE}: 0.2 \\
\ V2306 Cyg        & 2451762.9527(1)+0.181195(339)E  &  [9]   & 2451762.9527& {\em ASCA}: undef. \\
\ FO Aqr           & 2444782.869(2)+0.2020596(1)E    &  [10]  & 2444782.869 & {\em ASCA}: 0.011 ; {\em RXTE}: 0.014 \\
\ PQ Gem           & 2449333.984(4)+0.216359(3)E     &  [11]  & 2449333.984 & {\em ASCA}(1): 0.02 ; {\em ASCA}(2): 0.13 ; {\em RXTE}: 0.07\\
\ V709 Cas         & 2451048.0575(2)+0.2225(2)E      &  [12]  & 2451048.168 & {\em RXTE}: undef. \\
\ TV Col           & 2448267.4893(5)+0.2285999(2)E   &  [13]  & 2448267.4893& {\em ASCA}: 0.005 ; {\em RXTE}: 0.007 \\
\ TX Col$^{*}$     & 0.23825                         &  [14]  & --          & --\\
\ AE Aqr           & 2449281.42222(8)+0.41165553(5)E &  [15]  & 2449281.42222& {\em ASCA}: 0.0002 \\
\ V1062 Tau$^{*}$  & 0.41467                         &  [16]  & --          & -- \\
 \hline
\end{tabular}
$^{*}$ - No ephemeris available. \\ 
$^{\dag}$ numbers in brackets in the ephemerides indicate the uncertainty on the last digit. \\
References: [1] Buckley et al. (1998), [2] Hellier (1997), [3] Jablonski \& Steiner (1988), 
[4] Hellier \& Beardmore (2002), [5] Kaluzny \& Semeniuk (1988), [6] Haswell et al. (1997), 
[7] Taylor et al. (1999) ,[8] Thorstensen (1997), private communication, cited in Harlaftis \& Horne (1999), 
[9] Zharikov et al. (2002), [10] Patterson et al. (1998b), [11] Hellier (1997), 
[12] Bonnet-Bidaud et al. (2001), [13] Hellier (1993), [14] Mouchet et al. (1991), 
[15] \cite{CMMPH96}, [16] \cite{LLO04}
\end{table}

\clearpage

\begin{table}
\caption{Orbital modulation depths for {\em ASCA} data}
\begin{tabular}{lrrrrrr}
\hline 
 \multicolumn{1}{c}{Source}& 0.7--2~keV         & phase of  & $\chi^{2}_{r}$& 2--10~keV           & phase of & $\chi_{r}^{2}$\\
\                          &depth $\pm 1\sigma$ & min. flux & &depth $\pm 1\sigma$ & min. flux &\\
\hline 
\ {V1025 Cen}  &  3\%$\pm$3\%  &   0.73$\pm$0.17  &   5.8  &   2\%$\pm$2\%   &   0.76$\pm$0.16  &   4.7  \\
\ {BG CMi}     & 82\%$\pm$9\%  &   0.73$\pm$0.02  &    11  &  43\%$\pm$4\%   &   0.70$\pm$0.01  &   9.8  \\
\ {V1223 Sgr}  & 28\%$\pm$3\%  &   0.64$\pm$0.02  &    18  &  14\%$\pm$2\%   &   0.65$\pm$0.02  &    12  \\
\ {V2400 Oph}  &  4\%$\pm$2\%  &   0.65$\pm$0.09  &   8.2  &   4\%$\pm$2\%   &   0.57$\pm$0.10  &    15  \\
\ {AO Psc}     & 47\%$\pm$5\%  &   0.08$\pm$0.02  &    29  &  25\%$\pm$4\%   &   0.07$\pm$0.02  &    24  \\
\ {YY Dra}     &  9\%$\pm$4\%  &   0.69$\pm$0.07  &    10  &   5\%$\pm$5\%   &   0.70$\pm$0.14  &    15  \\ 
\ {LS Peg}     & 21\%$\pm$10\% &   0.09$\pm$0.09  &   1.5  &  21\%$\pm$8\%   &   0.05$\pm$0.06  &   2.2  \\
\ {V405 Aur(1)}&  6\%$\pm$2\%  &   0.71$\pm$0.07  &   4.0  &   2\%$\pm$1\%   &   0.68$\pm$0.15  &   1.7  \\
\ {V405 Aur(2)}& 21\%$\pm$4\%  &   0.54$\pm$0.03  &   6.3  &   9\%$\pm$3\%   &   0.57$\pm$0.04  &   3.4  \\
\ {V2306 Cyg}  & 25\%$\pm$8\%  &   0.56$\pm$0.06  &   3.3  &  18\%$\pm$6\%   &   0.56$\pm$0.04  &   3.2  \\ 
\ {FO Aqr}     &102\%$\pm$13\% &   0.04$\pm$0.02  &    26  &  70\%$\pm$10\%  &   0.99$\pm$0.02  &    90  \\
\ {PQ Gem(1)}  & 10\%$\pm$3\%  &   0.83$\pm$0.05  &   5.1  &   4\%$\pm$1\%   &   0.79$\pm$0.05  &   1.3  \\
\ {PQ Gem(2)}  &  3\%$\pm$4\%  &   0.62$\pm$0.26  &   6.2  &   3\%$\pm$3\%   &   0.46$\pm$0.14  &   4.6  \\
\ {TV Col}     & 47\%$\pm$6\%  &   0.98$\pm$0.02  &    32  &  28\%$\pm$4\%   &   0.98$\pm$0.02  &    22  \\
\ {TX Col}     & 58\%$\pm$6\%  &   0.20$\pm$0.01  &   7.0  &  49\%$\pm$3\%   &   0.22$\pm$0.01  &   3.9  \\
\ {AE Aqr}     & 34\%$\pm$9\%  &   0.92$\pm$0.03  &    14  &  30\%$\pm$7\%   &   0.89$\pm$0.04  &   3.1  \\
\ {V1062 Tau}  & 35\%$\pm$12\% &   0.16$\pm$0.06  &    32  &  17\%$\pm$8\%   &   0.16$\pm$0.08  &    40  \\
\hline
\end{tabular}
\end{table}

\clearpage

\begin{table}
\caption{Orbital modulation depths for {\em RXTE} data}

\begin{tabular}{lrrrrrr}
\hline
 \multicolumn{1}{c}{Source}& 2--4~keV  & phase of & $\chi^{2}_{r}$ & 4--6~keV& phase of & $\chi^{2}_{r}$\\
\     & depth $\pm 1\sigma$ & min. flux& & depth $\pm 1\sigma$ & min. flux & \\
\hline
\ {V1025 Cen}   & 13\%$\pm$3\%    &  0.53$\pm$0.03   &   1.5  &  14\%$\pm$3\%    &  0.61$\pm$0.03   &   2.7   \\
\ {BG CMi}      & 39\%$\pm$9\%    &  0.72$\pm$0.02   &    29  &  32\%$\pm$5\%    &  0.70$\pm$0.02   &    31   \\
\ {V1223 Sgr}   &  7\%$\pm$2\%    &  0.66$\pm$0.04   &    26  &   8\%$\pm$2\%    &  0.65$\pm$0.04   &    88   \\
\ {V2400 Oph}   &  9\%$\pm$2\%    &  0.72$\pm$0.02   &   8.2  &  10\%$\pm$2\%    &  0.71$\pm$0.02   &    23   \\
\ {AO Psc}      &  7\%$\pm$2\%    &  0.10$\pm$0.05   &   9.4  &   3\%$\pm$1\%    &  0.05$\pm$0.07   &    10   \\
\ {YY Dra}      &  5\%$\pm$3\%    &  0.88$\pm$0.09   &   1.4  &   1\%$\pm$2\%    &  undefined       &   2.8   \\
\ {V405 Aur}    &  8\%$\pm$2\%    &  0.69$\pm$0.04   &   1.2  &   7\%$\pm$1\%    &  0.74$\pm$0.03   &   1.3   \\
\ {FO Aqr}      & 57\%$\pm$14\%   &  0.81$\pm$0.03   &   300  &  44\%$\pm$8\%    &  0.78$\pm$0.03   &   770   \\
\ {PQ Gem}      &  6\%$\pm$3\%    &  0.73$\pm$0.07   &   4.5  &   6\%$\pm$2\%    &  0.81$\pm$0.05   &   5.0   \\
\ {V709 Cas}    &  3\%$\pm$2\%    &  0.14$\pm$0.10   &   5.8  &  3\% $\pm$1\%    &  0.13$\pm$0.06   &   5.5   \\
\ {TV Col}      & 25\%$\pm$4\%    &  0.88$\pm$0.03   &    61  &  12\%$\pm$3\%    &  0.89$\pm$0.04   &    68   \\
\ {TX Col}      & 43\%$\pm$4\%    &  0.86$\pm$0.01   &    12  &  37\%$\pm$2\%    &  0.87$\pm$0.01   &   8.9   \\
\ {V1062 Tau}   & 25\%$\pm$11\%   &  0.31$\pm$0.06   &    80  &  20\%$\pm$7\%    &  0.28$\pm$0.05   &   120   \\
\hline
\end{tabular}

\end{table}

\begin{table}

\begin{tabular}{lrrrrrr}
\hline
 \multicolumn{1}{c}{Source} & 6--10~keV & phase of &  $\chi^{2}_{r}$& 10-20~keV & phase of  &  $\chi^{2}_{r}$ \\
\                           & depth $\pm 1\sigma$ & min. flux  &  & depth $\pm 1\sigma$ & min. flux &   \\
\hline
\ {V1025 Cen}   & 12\%$\pm$3\%  &  0.52$\pm$0.04  &   2.1  &   31\%$\pm$6\%  &   0.58$\pm$0.03  &   1.5  \\
\ {BG CMi}      & 31\%$\pm$4\%  &  0.70$\pm$0.01  &    22  &   34\%$\pm$4\%  &   0.68$\pm$0.01  &   5.6  \\
\ {V1223 Sgr}   &  9\%$\pm$2\%  &  0.65$\pm$0.03  &   110  &   12\%$\pm$3\%  &   0.62$\pm$0.03  &    70  \\
\ {V2400 Oph}   & 10\%$\pm$3\%  &  0.69$\pm$0.03  &    39  &   11\%$\pm$4\%  &   0.69$\pm$0.04  &   9.2  \\
\ {AO Psc}      &  2\%$\pm$1\%  &  0.98$\pm$0.13  &   9.1  &    3\%$\pm$1\%  &   0.08$\pm$0.05  &   0.7  \\
\ {YY Dra}      &  1\%$\pm$2\%  &  undefined      &   2.2  &    2\%$\pm$3\%  &   undefined      &   0.8  \\
\ {V405 Aur}    &  7\%$\pm$1\%  &  0.77$\pm$0.03  &   1.1  &   10\%$\pm$2\%  &   0.85$\pm$0.04  &   0.7  \\
\ {FO Aqr}      & 34\%$\pm$6\%  &  0.78$\pm$0.03  &   650  &   26\%$\pm$6\%  &   0.81$\pm$0.03  &   130  \\
\ {PQ Gem}      &  6\%$\pm$2\%  &  0.83$\pm$0.05  &   6.1  &    4\%$\pm$2\%  &   0.75$\pm$0.10  &   1.7  \\
\ {V709 Cas}    &  2\%$\pm$1\%  &  0.09$\pm$0.10  &   6.3  &    2\%$\pm$1\%  &   0.94$\pm$0.10  &   2.4  \\
\ {TV Col}      &  6\%$\pm$2\%  &  0.94$\pm$0.06  &    57  &    7\%$\pm$3\%  &   0.13$\pm$0.07  &   7.7  \\
\ {TX Col}      & 28\%$\pm$3\%  &  0.87$\pm$0.02  &    11  &   28\%$\pm$4\%  &   0.90$\pm$0.02  &   5.2  \\
\ {V1062 Tau}   & 18\%$\pm$7\%  &  0.26$\pm$0.05  &   160  &   25\%$\pm$8\%  &   0.24$\pm$0.04  &    71  \\
\hline
\end{tabular}

\end{table}

\clearpage

\begin{figure}[h]
\caption{Upper panels: phase folded {\em ASCA} lightcurves of each object
with the best fit sinusoid overplotted in each case. Lower panels: power spectra
of the {\em ASCA} lightcurves of each object. The low frequency range is shown
on an expanded scale in each case to aid visibility of the orbital components;
 $\omega$ - indicates the spin frequency, $\Omega$ - indicates the orbital 
frequency and the horizontal dashed line indicates the noise level in the power 
spectra, for details see text. (Nb. For the full set of objects, see the on-line
version of this paper.}
\end{figure}  

\clearpage

\begin{figure}[h]
\epsfig{file=2887f1a1, angle=270, width=14cm}
\end{figure}  

\begin{figure}[h]
\epsfig{file=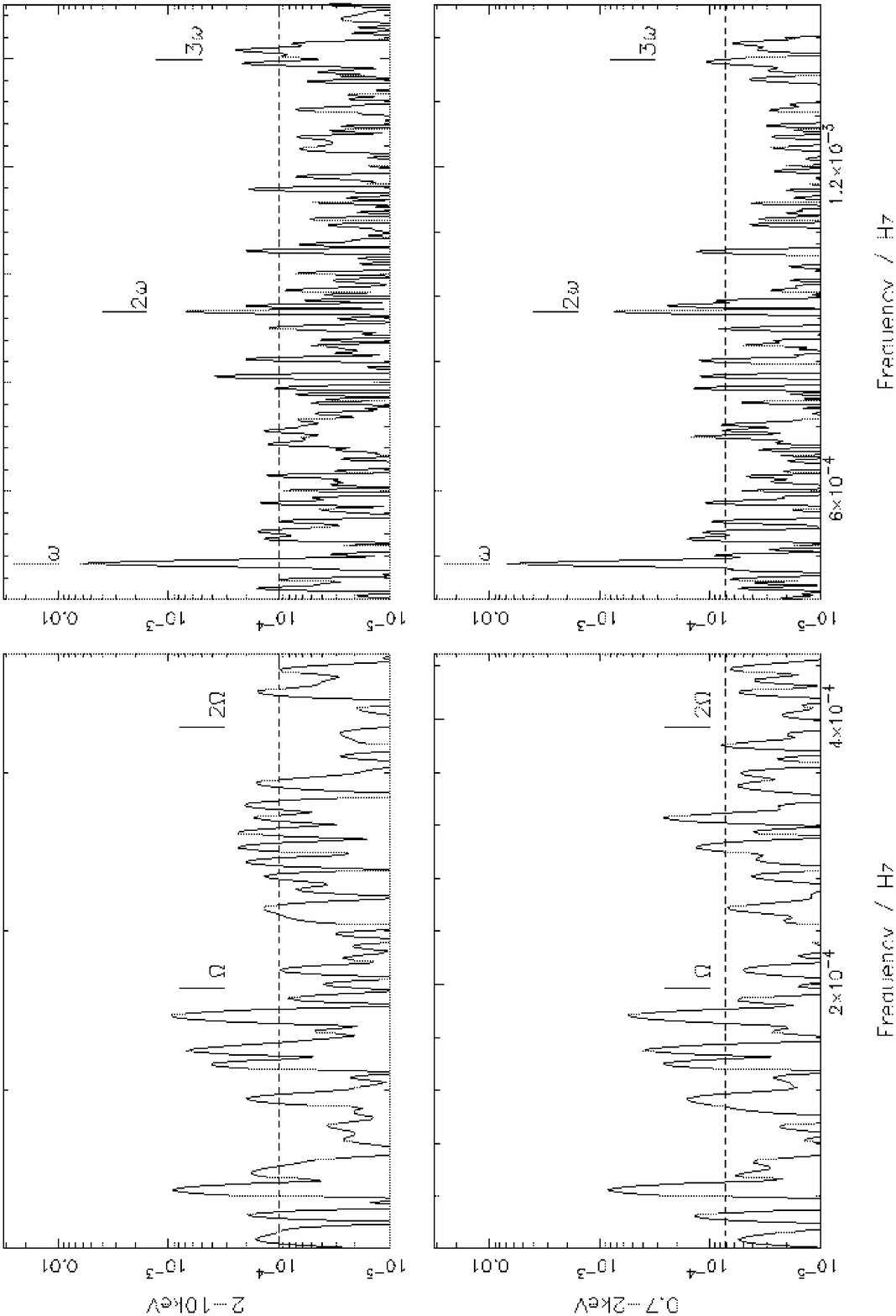, angle=270, width=14cm}
\setcounter{figure}{0}
\caption{(a) V1025 Cen}
\end{figure}  

\clearpage

\begin{figure}[h]
\epsfig{file=2887f1b1.eps, angle=270, width=14cm}
\end{figure}

\begin{figure}[h]
\epsfig{file=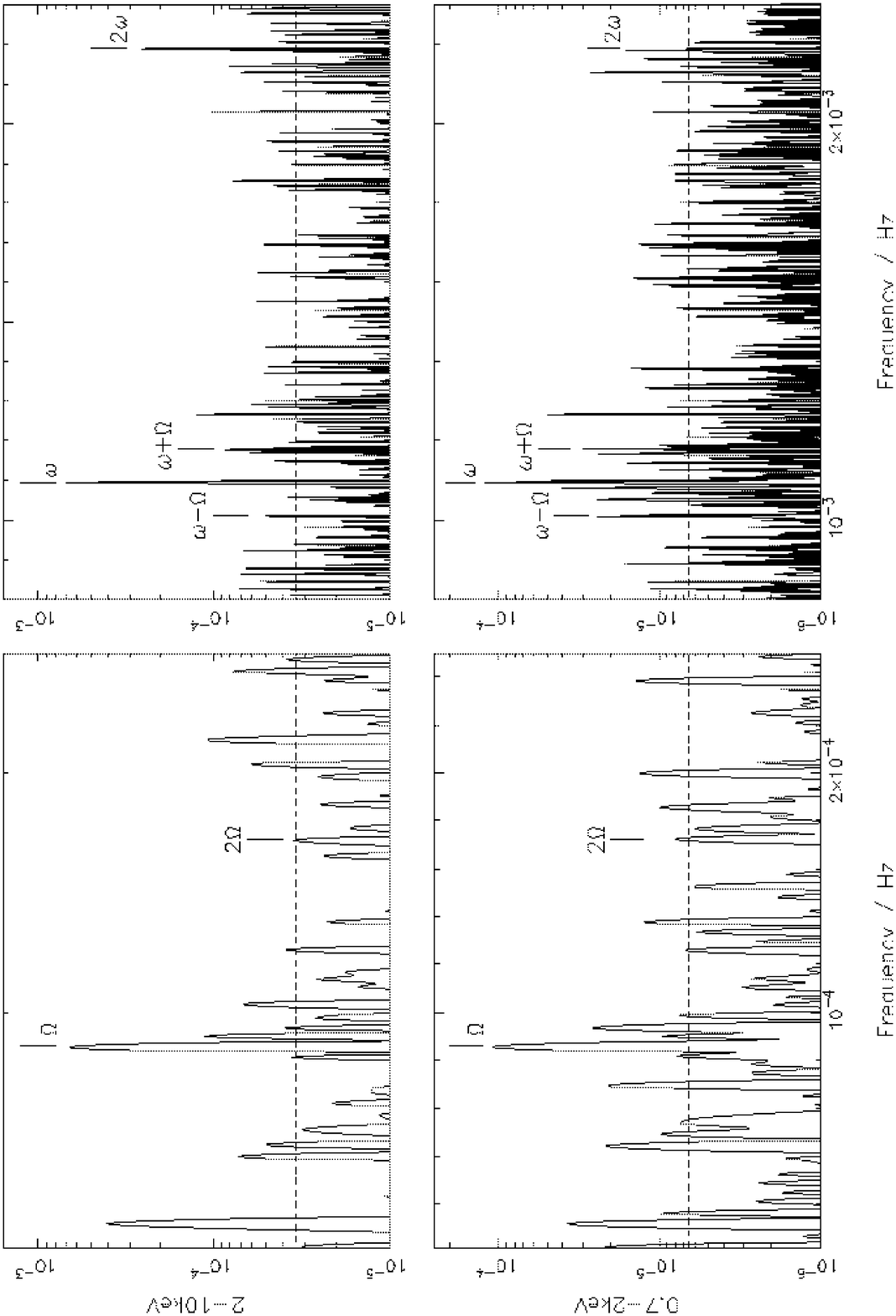, angle=270, width=14cm} 
\setcounter{figure}{0}
\caption{(b) BG CMi}
\end{figure}

\clearpage

\begin{figure}[h]
\epsfig{file=2887f1c1.eps, angle=270, width=14cm}
\end{figure}

\begin{figure}[h]
\epsfig{file=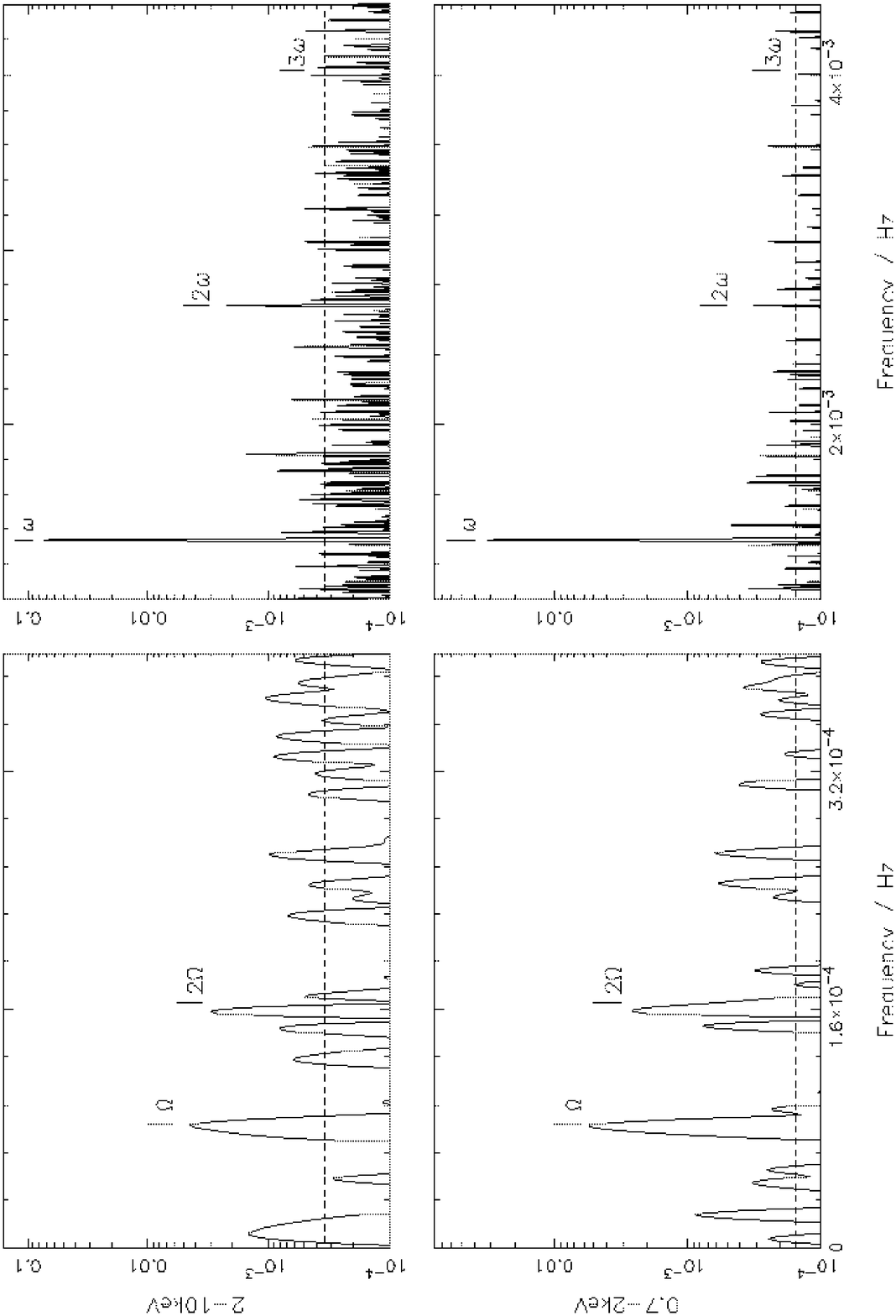, angle=270, width=14cm} 
\setcounter{figure}{0}
\caption{(c) V1223 Sgr}
\end{figure}

\clearpage

\begin{figure}[h]
\epsfig{file=2887f1d1.eps, angle=270, width=14cm}
\end{figure}  

\begin{figure}[h]
\epsfig{file=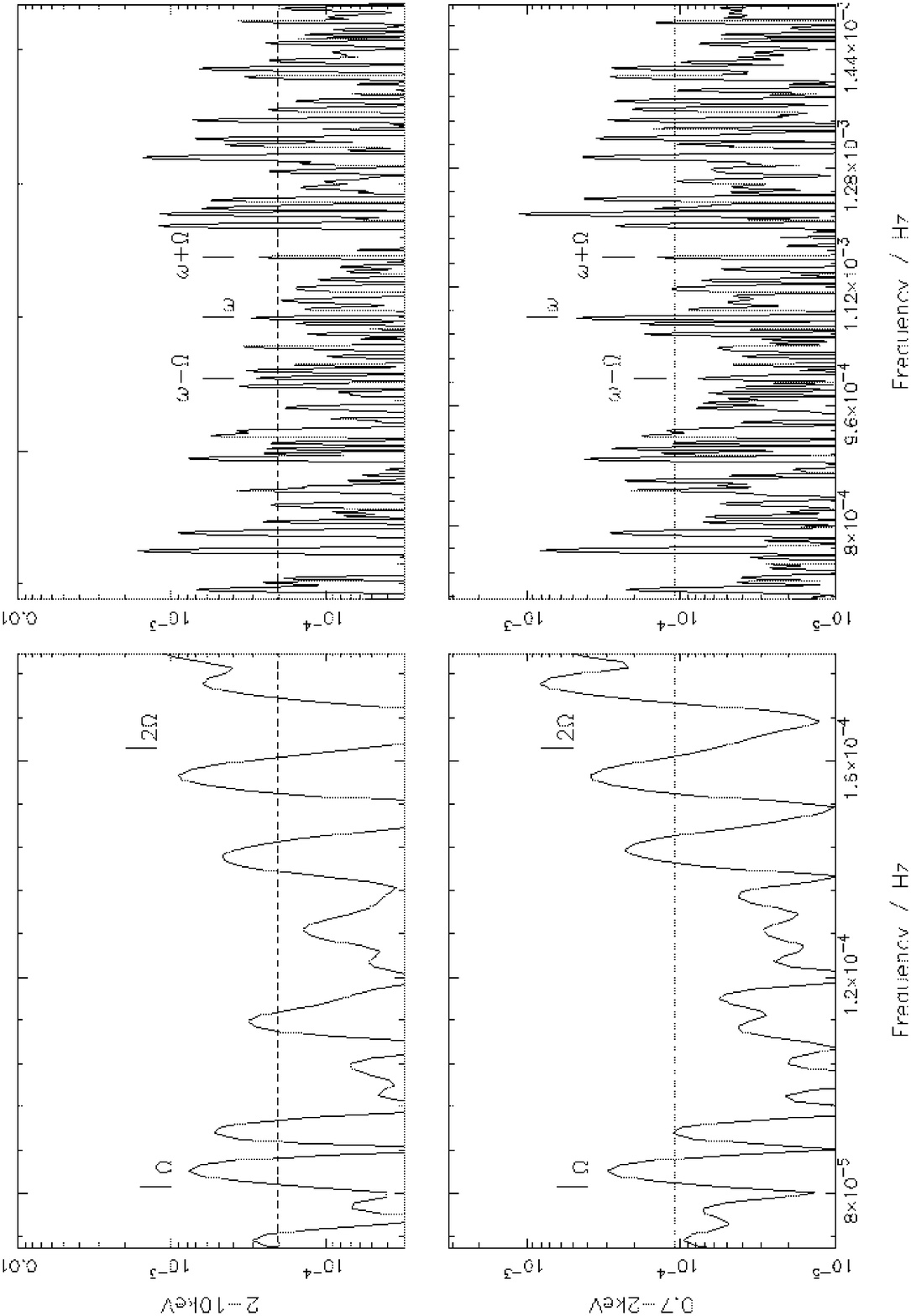, angle=270, width=14cm} 
\setcounter{figure}{0}
\caption{(d) V2400 Oph}
\end{figure}  

\clearpage

\begin{figure}[h]
\epsfig{file=2887f1e1.eps, angle=270, width=14cm}
\end{figure}  

\begin{figure}[h]
\epsfig{file=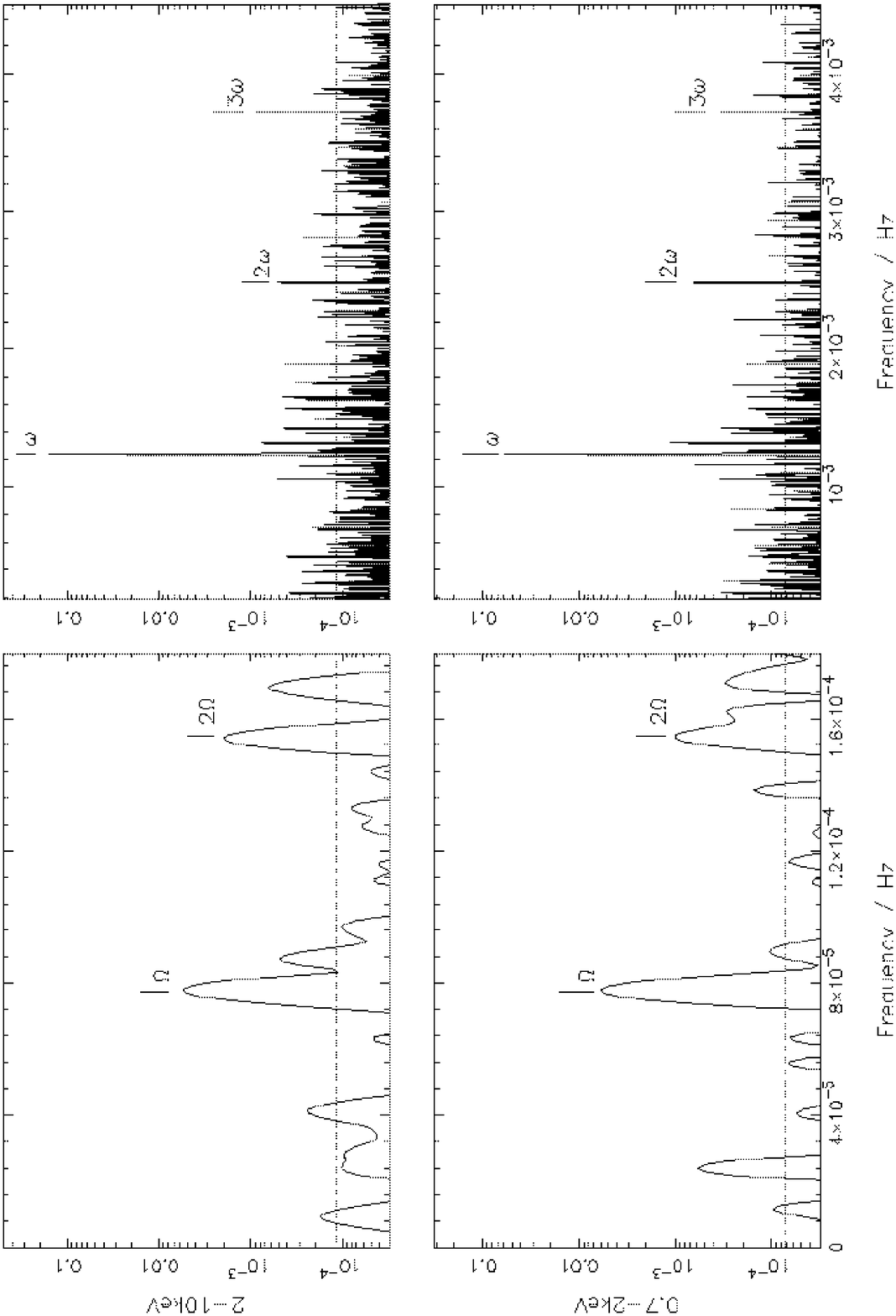, angle=270, width=14cm} 
\setcounter{figure}{0}
\caption{(e) AO Psc}
\end{figure}  

\clearpage

\begin{figure}[h]
\epsfig{file=2887f1f1.eps, angle=270, width=14cm}
\end{figure}  

\begin{figure}[h]
\epsfig{file=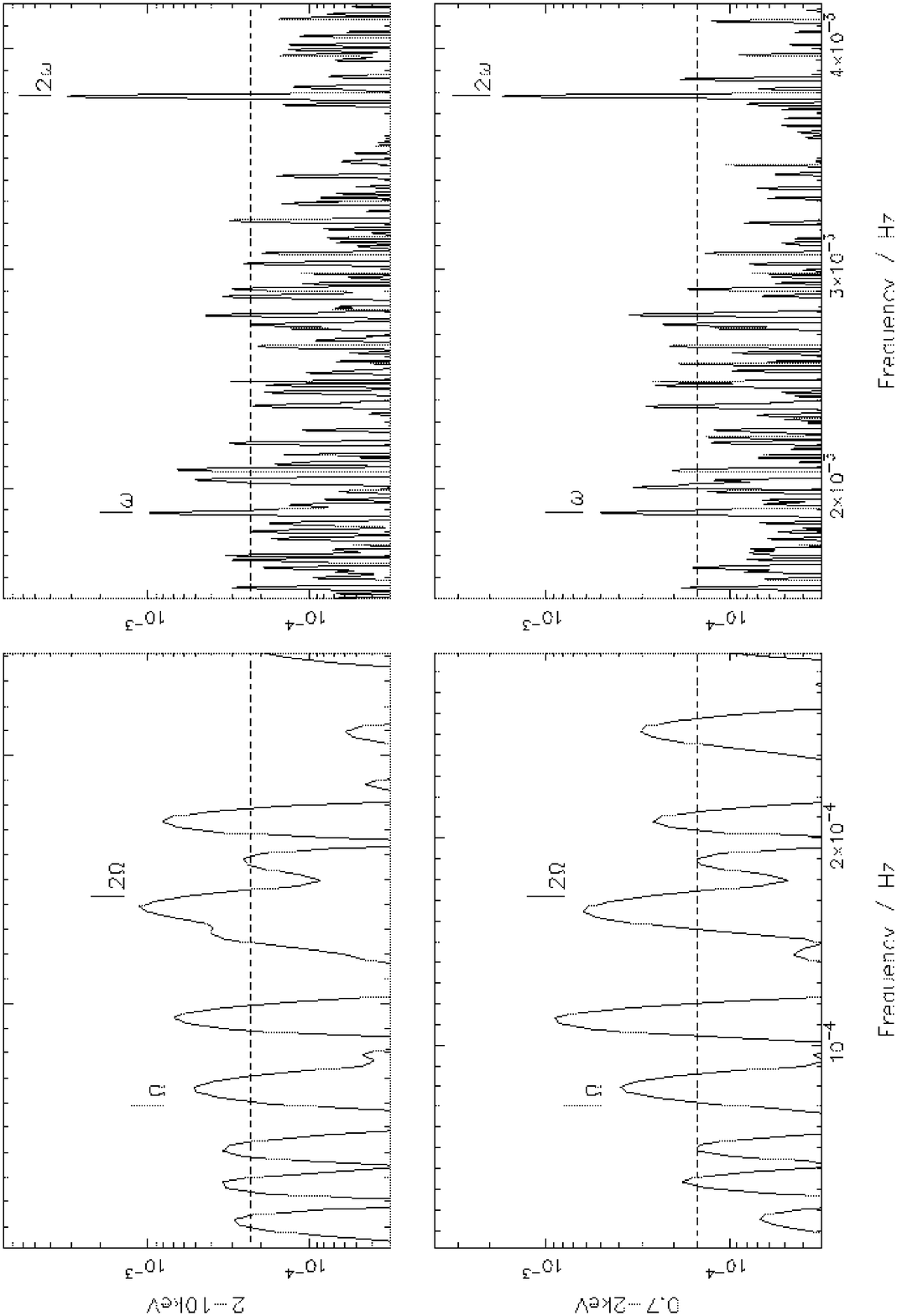, angle=270, width=14cm} 
\setcounter{figure}{0}
\caption{(f) YY Dra}
\end{figure}  

\clearpage

\begin{figure}[h]
\epsfig{file=2887f1g1.eps, angle=270, width=14cm}
\end{figure}  

\begin{figure}[h]
\epsfig{file=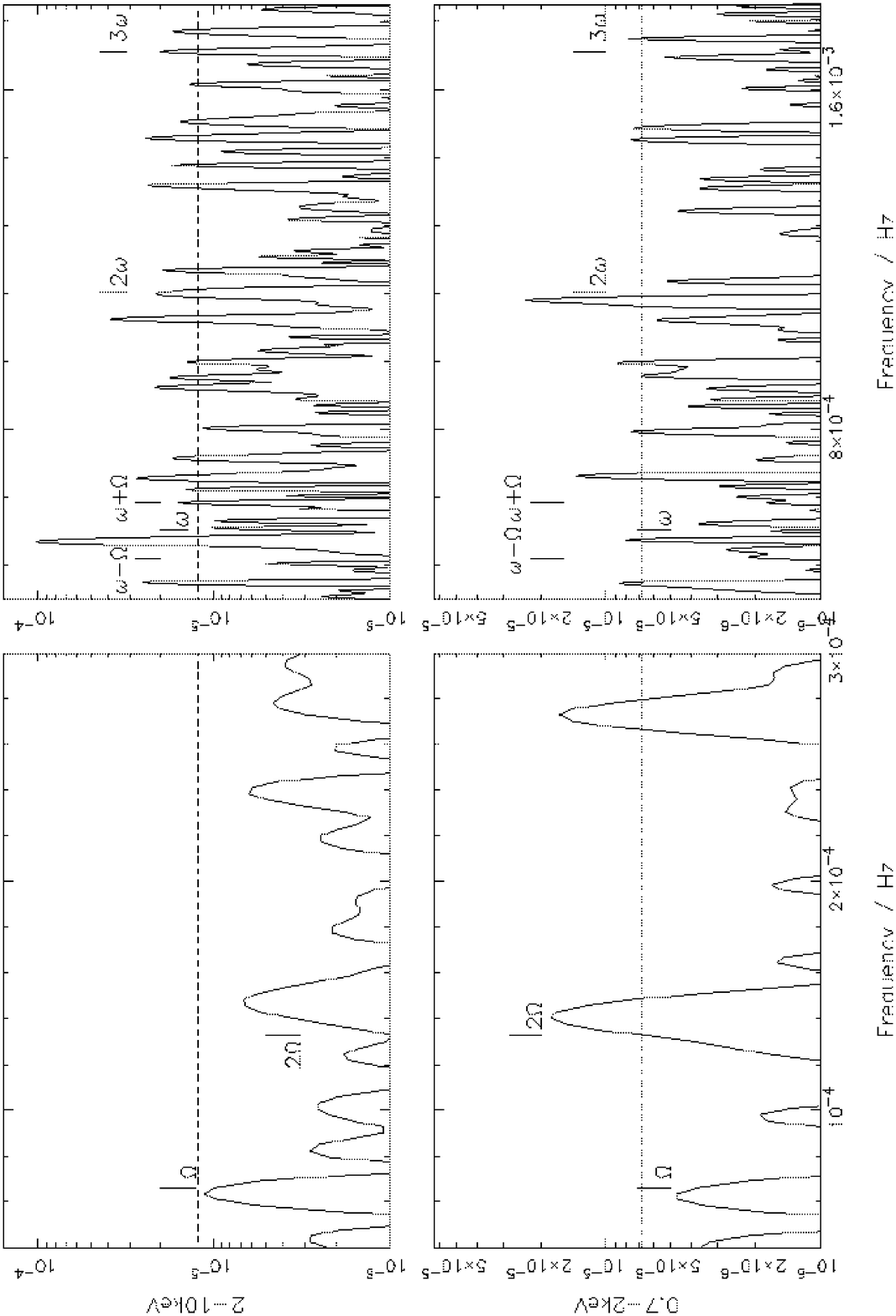, angle=270, width=14cm} 
\setcounter{figure}{0}
\caption{(g) LS Peg}
\end{figure}  

\clearpage

\begin{figure}[h]
\epsfig{file=2887f1h1.eps, angle=270, width=14cm}
\end{figure}  

\begin{figure}[h]
\epsfig{file=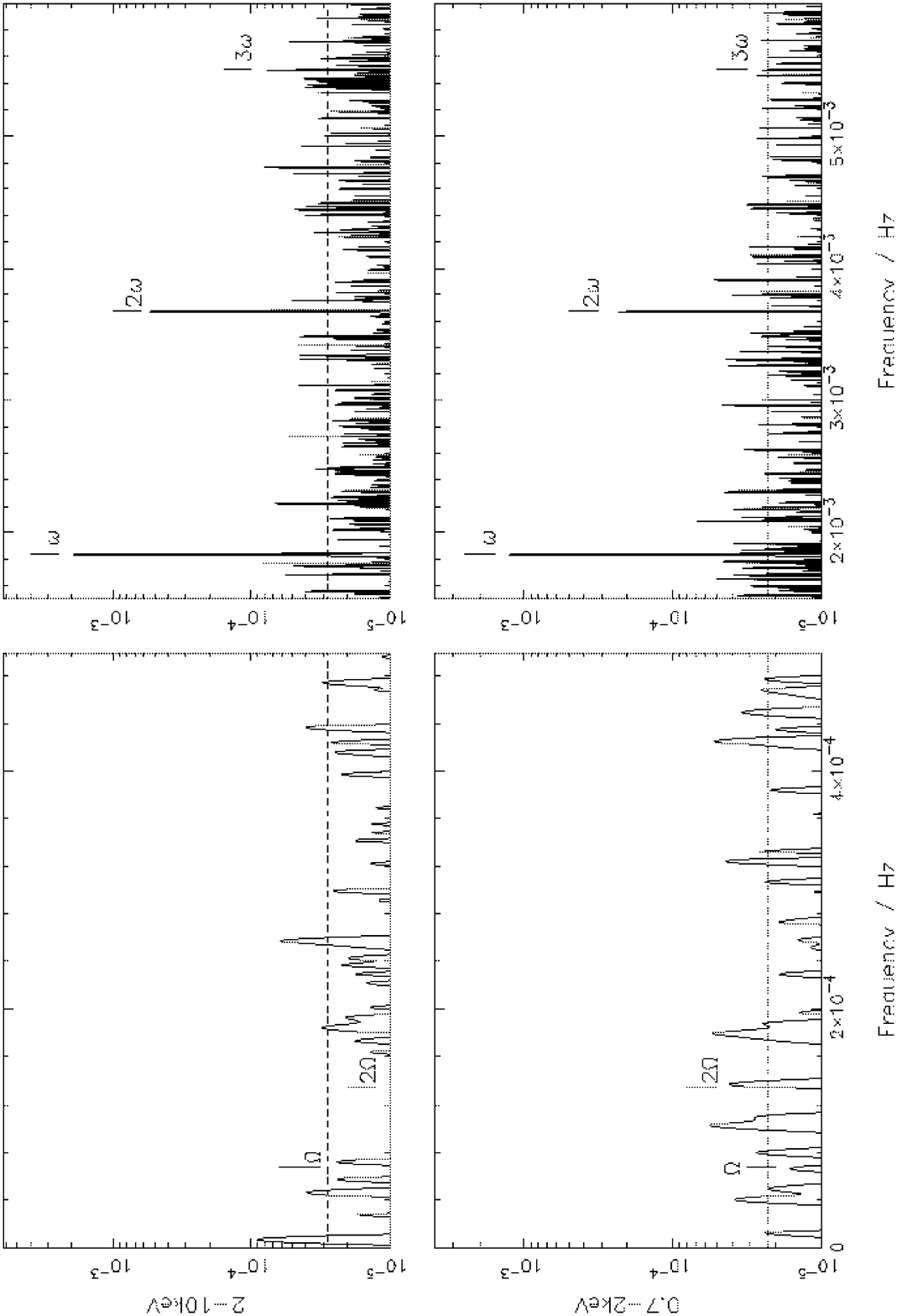, angle=270, width=14cm}
\setcounter{figure}{0}
\caption{(h) V405 Aur (1)} 
\end{figure}  

\clearpage

\begin{figure}[h]
\epsfig{file=2887f1i1.eps, angle=270, width=14cm}
\end{figure}  

\begin{figure}[h]
\epsfig{file=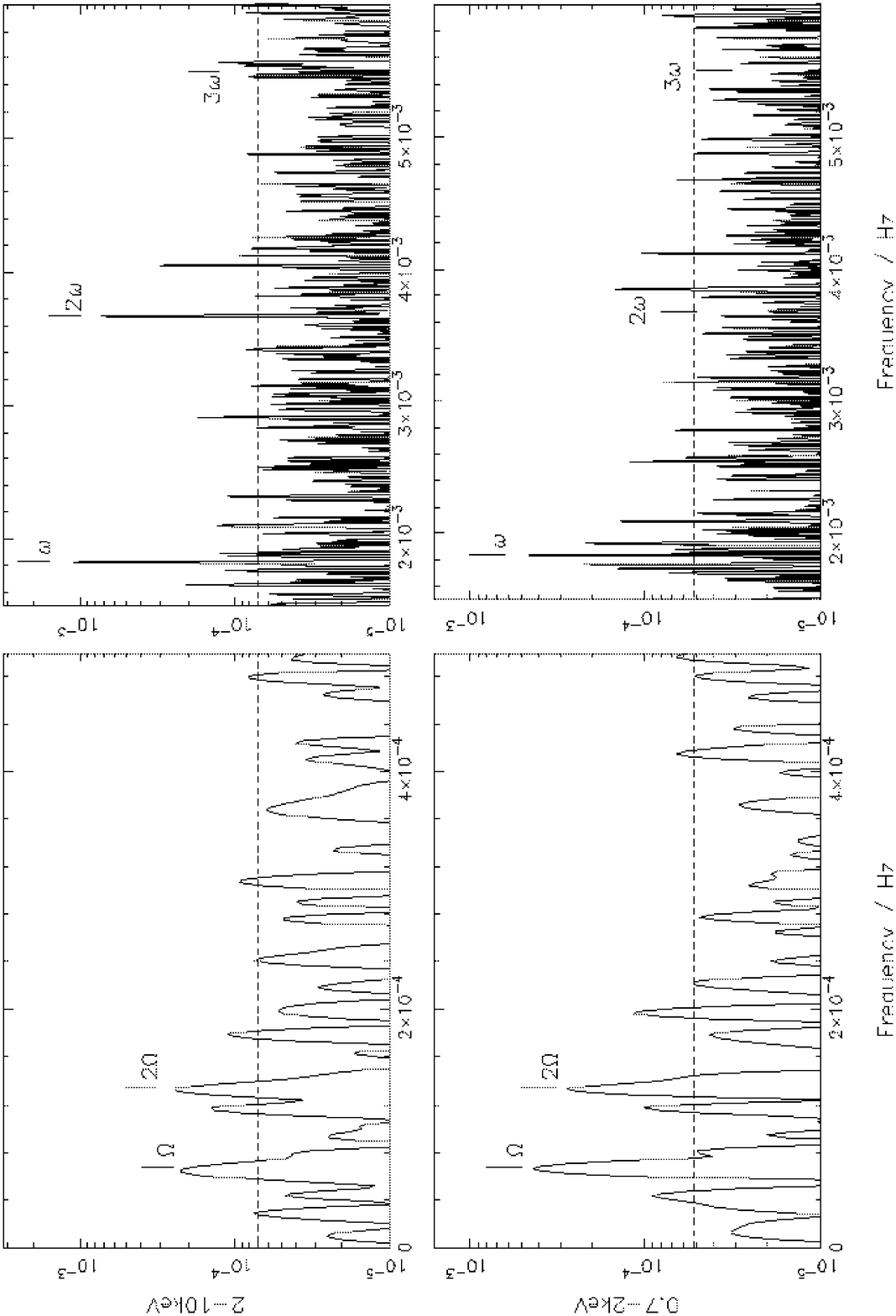, angle=270, width=14cm}
\setcounter{figure}{0}
\caption{(i) V405 Aur (2)} 
\end{figure}  

\clearpage

\begin{figure}[h]
\epsfig{file=2887f1j1.eps, angle=270, width=14cm}
\end{figure}  

\begin{figure}[h]
\epsfig{file=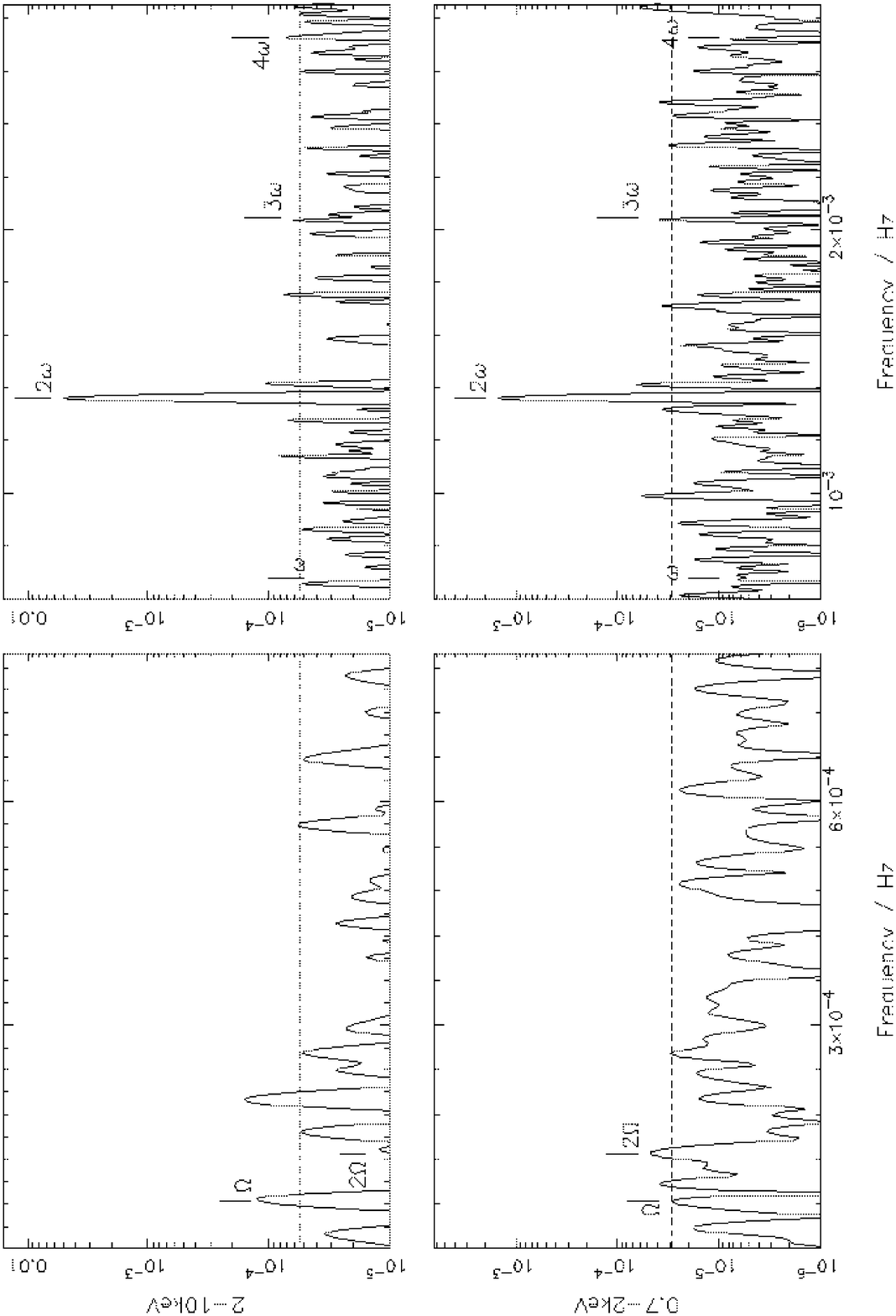, angle=270, width=14cm}
\setcounter{figure}{0}
\caption{(j) V2306 Cyg} 
\end{figure}  

\clearpage

\begin{figure}[h]
\epsfig{file=2887f1k1.eps, angle=270, width=14cm}
\end{figure}  

\begin{figure}[h]
\epsfig{file=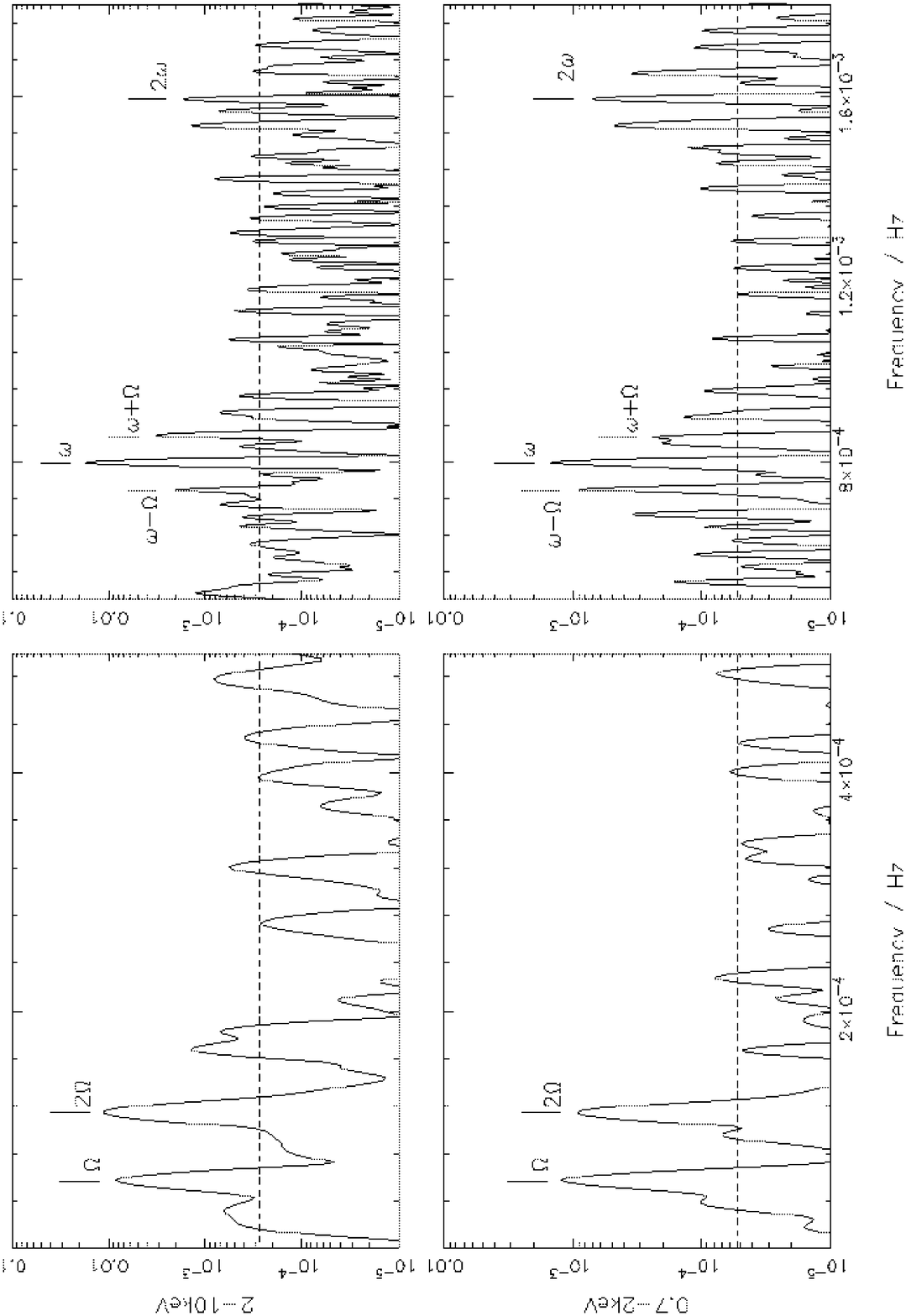, angle=270, width=14cm}
\setcounter{figure}{0}
\caption{(k) FO Aqr} 
\end{figure}  

\clearpage

\begin{figure}[h]
\epsfig{file=2887f1l1.eps, angle=270, width=14cm}
\end{figure}  

\begin{figure}[h]
\epsfig{file=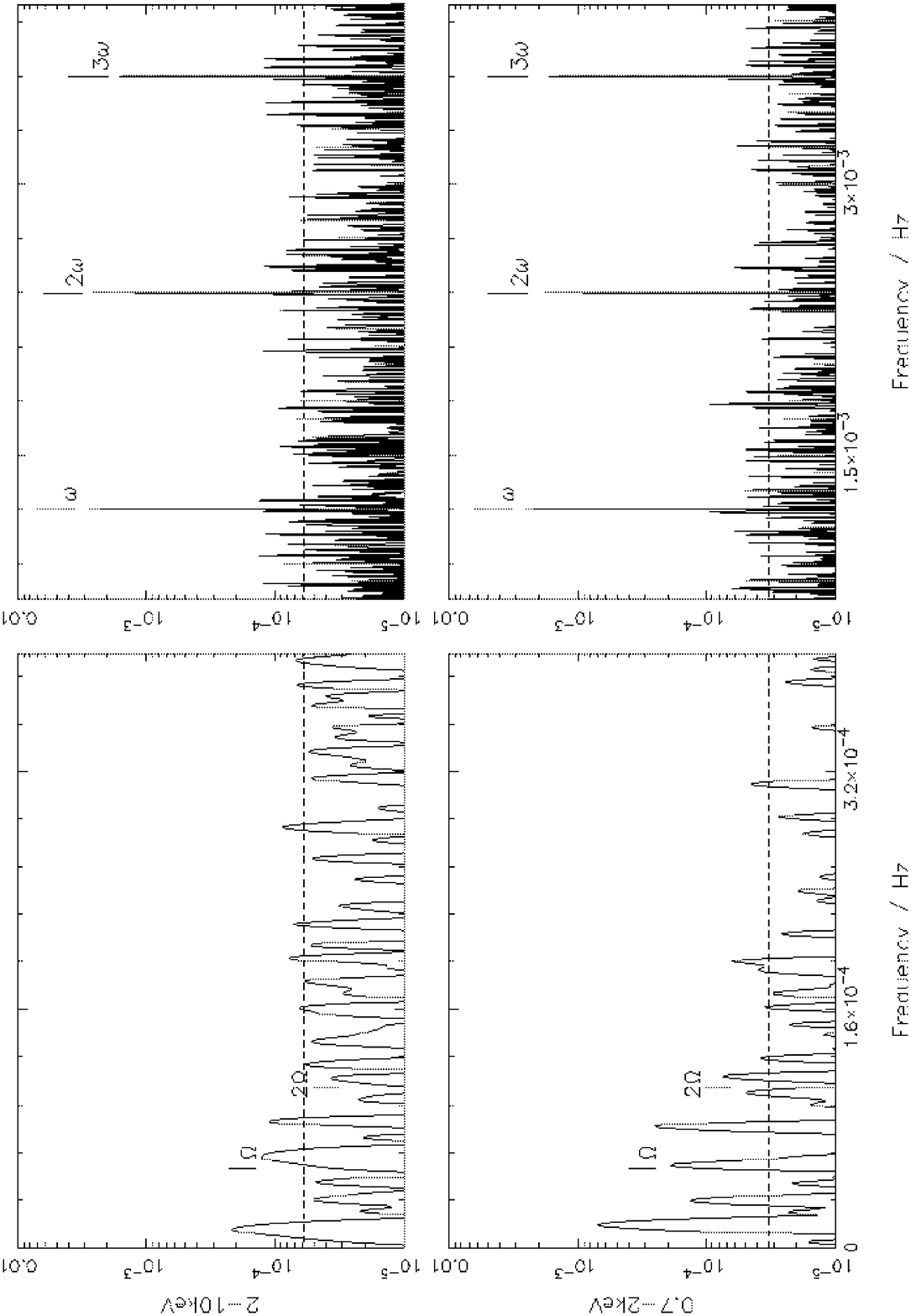, angle=270, width=14cm} 
\setcounter{figure}{0}
\caption{(l) PQ Gem (1)}
\end{figure}  

\clearpage

\begin{figure}[h]
\epsfig{file=2887f1m1.eps, angle=270, width=14cm}
\end{figure}  

\begin{figure}[h]
\epsfig{file=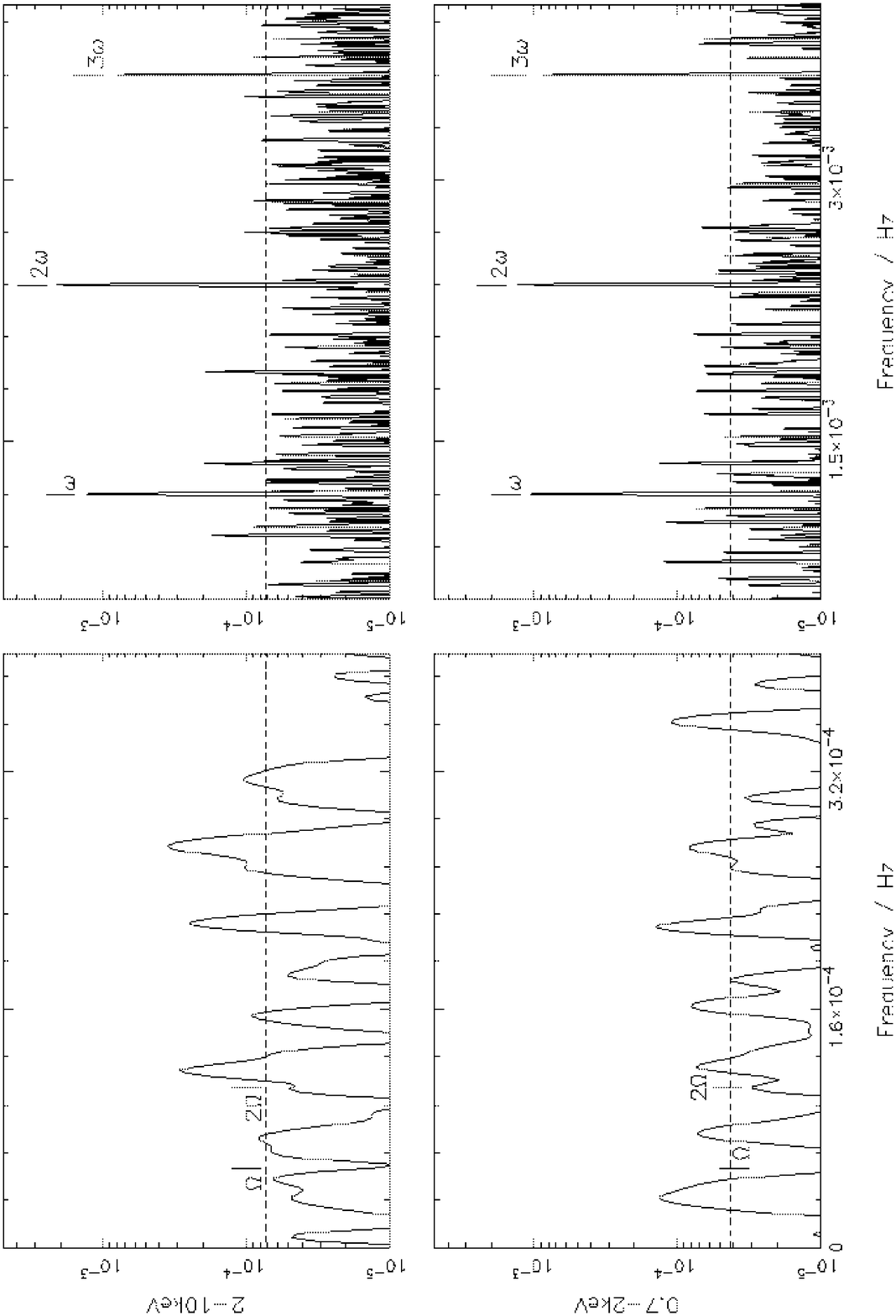, angle=270, width=14cm} 
\setcounter{figure}{0}
\caption{(m) PQ Gem (2)}
\end{figure}  

\clearpage

\begin{figure}[h]
\epsfig{file=2887f1n1.eps, angle=270, width=14cm}
\end{figure}  

\begin{figure}[h]
\epsfig{file=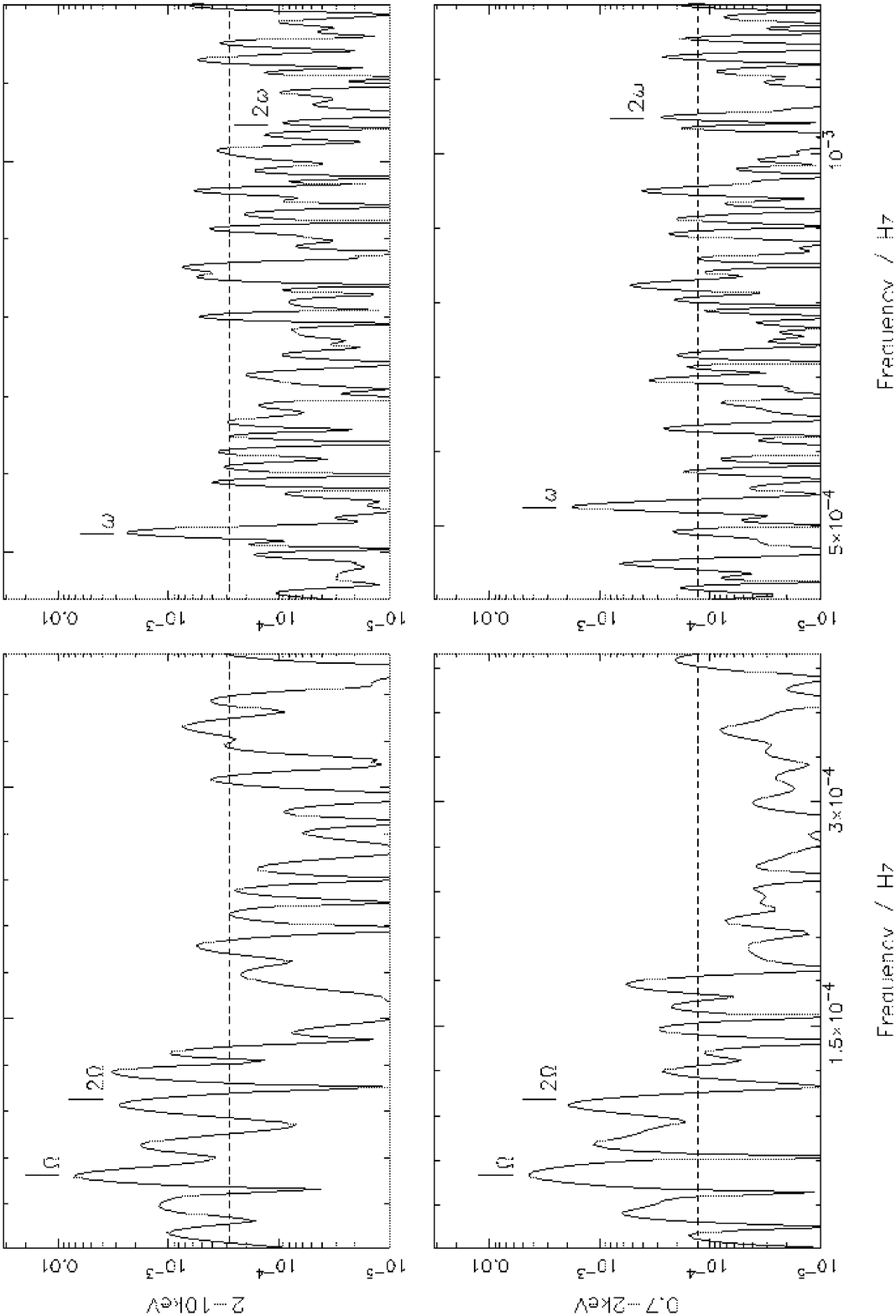, angle=270, width=14cm}
\setcounter{figure}{0}
\caption{(n) TV Col} 
\end{figure}  

\clearpage

\begin{figure}[h]
\epsfig{file=2887f1o1.eps, angle=270, width=14cm}
\end{figure}  

\begin{figure}[h]
\epsfig{file=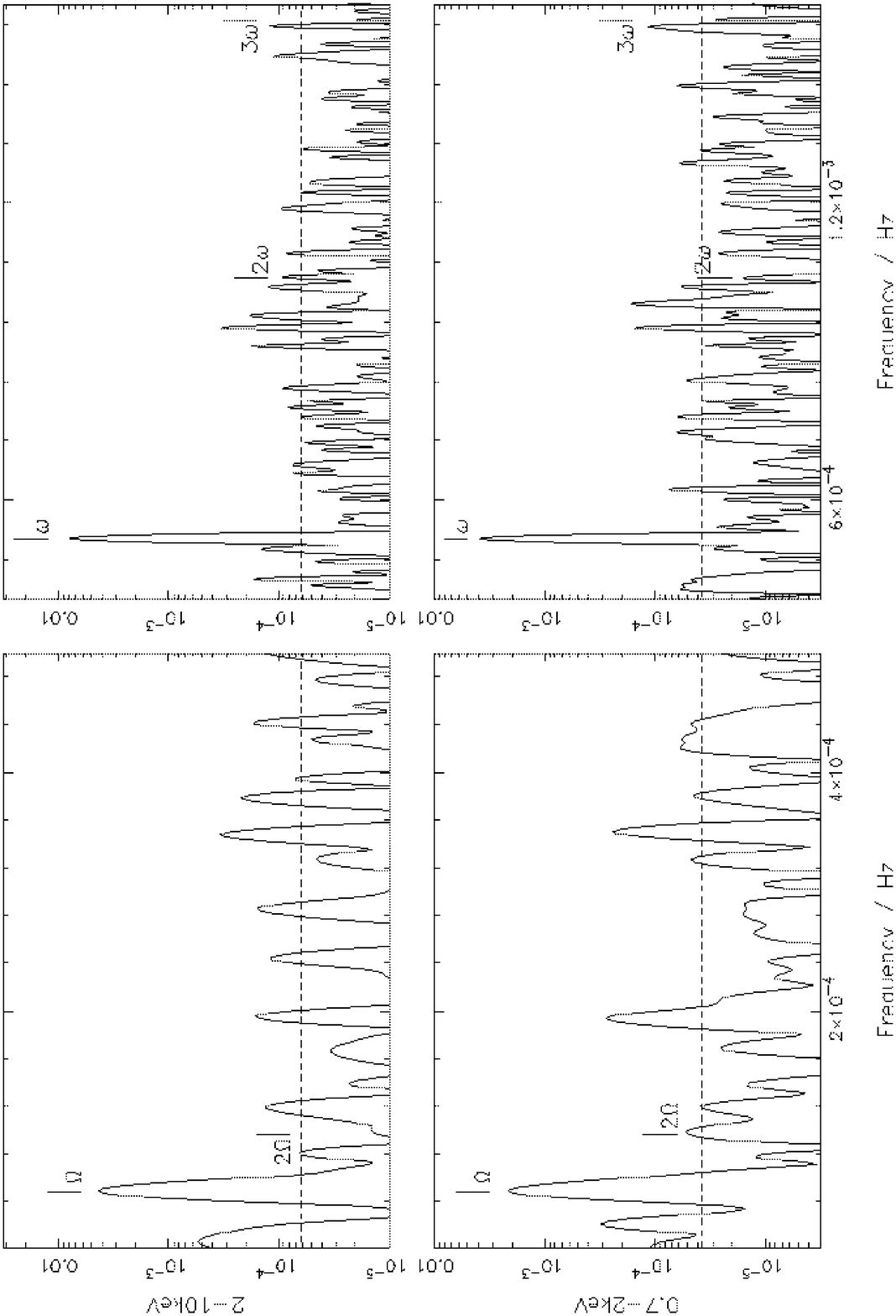, angle=270, width=14cm}
\setcounter{figure}{0}
\caption{(o) TX Col} 
\end{figure}  

\clearpage

\begin{figure}[h]
\epsfig{file=2887f1p1.eps, angle=270, width=14cm}
\end{figure}  

\begin{figure}[h]
\epsfig{file=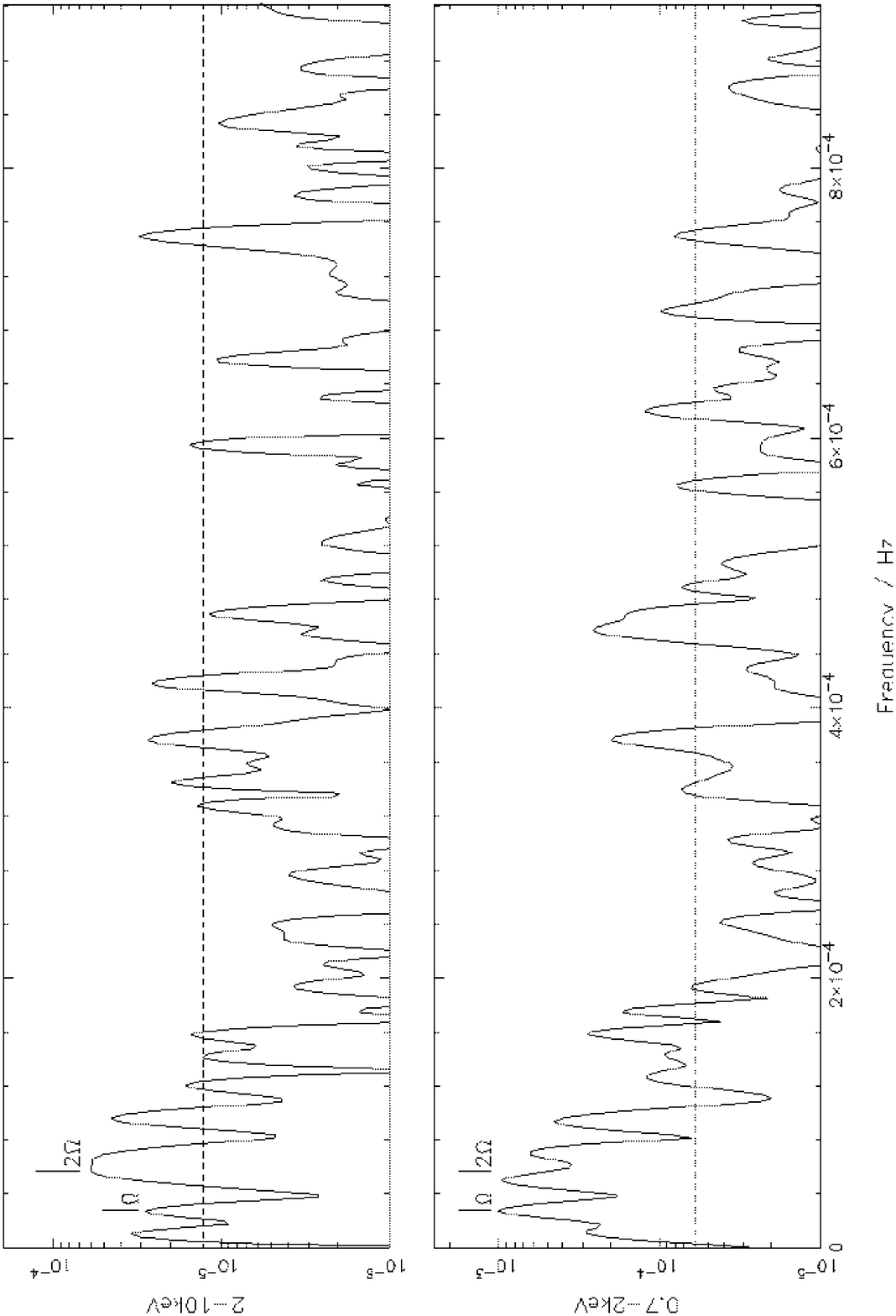, angle=270, width=14cm} 
\setcounter{figure}{0}
\caption{(p) AE Aqr}
\end{figure}  

\clearpage

\begin{figure}[h]
\epsfig{file=2887f1q1.eps, angle=270, width=14cm}
\end{figure}  

\begin{figure}[h]
\epsfig{file=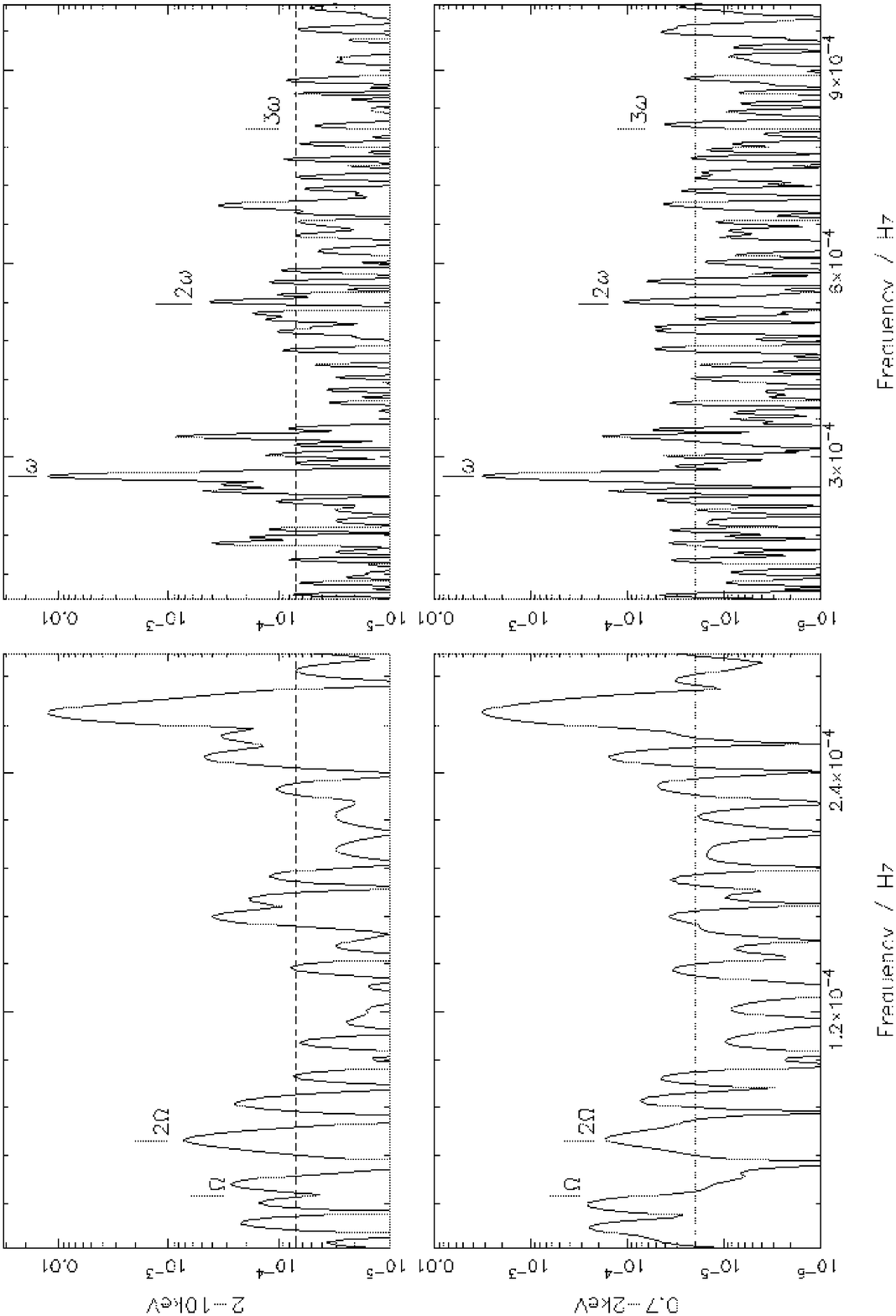, angle=270, width=14cm} 
\setcounter{figure}{0}
\caption{(q) V1062 Tau}
\end{figure}

\clearpage

\begin{figure}[h]
\caption{Upper panels: phase folded {\em RXTE} lightcurves of each object
with the best fit sinusoid overplotted in each case. Lower panels: power spectra
of the {\em RXTE} lightcurves of each object. The low frequency range is shown
on an expanded scale in each case to aid visibility of the orbital components;
 $\omega$ - indicates the spin frequency, $\Omega$ - indicates the orbital 
frequency and the horizontal dashed line indicates the noise level in the power 
spectra, for details see text. (Nb. For the full set of objects, see the on-line
version of this paper.}
\end{figure}

\clearpage

\begin{figure}[h]
\epsfig{file=2887f2a1.eps, angle=270, width=14cm} 
\end{figure}  

\begin{figure}[h]
\epsfig{file=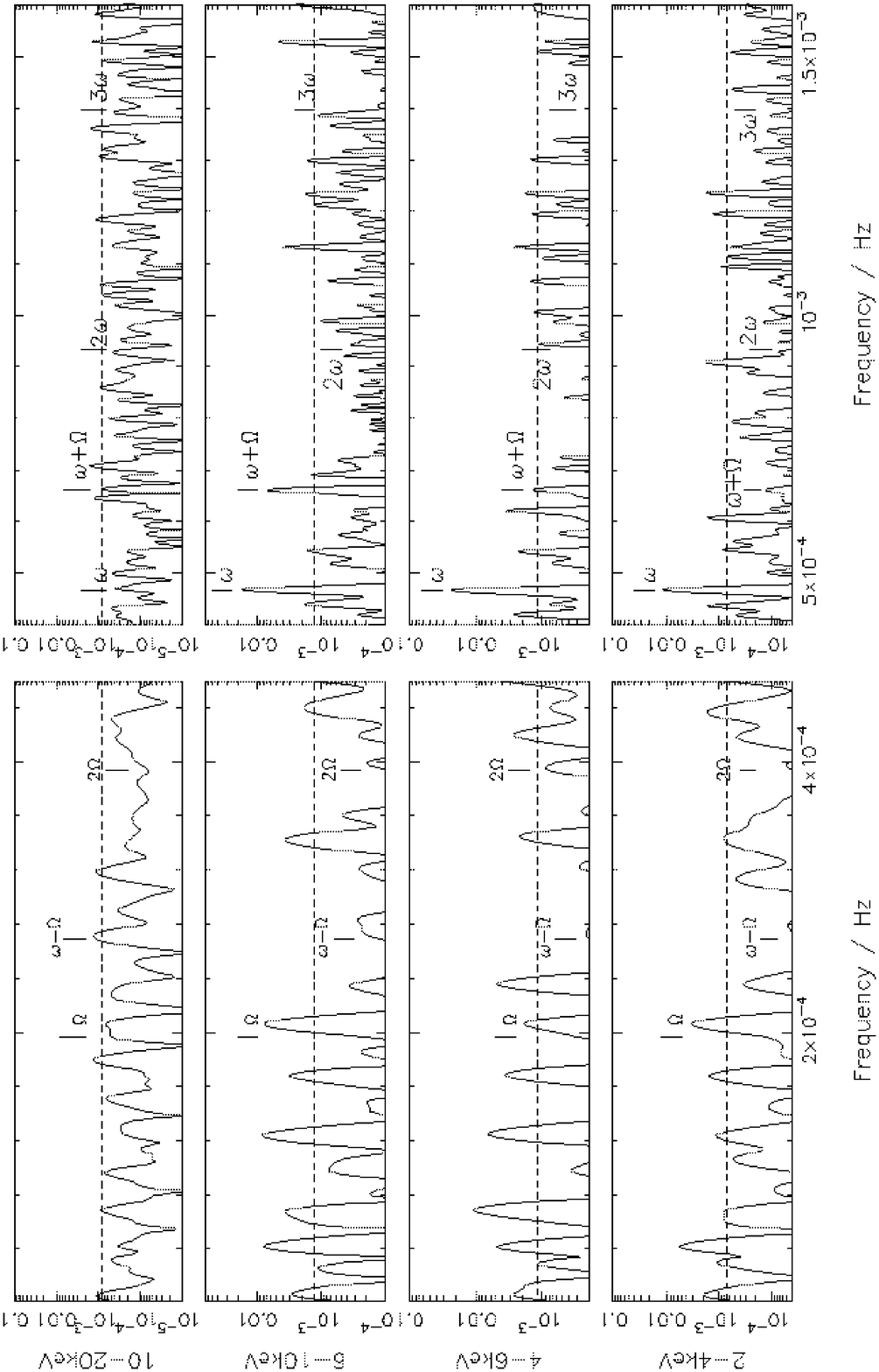, angle=270, width=14cm} 
\setcounter{figure}{1}
\caption{(a) V1025 Cen}
\end{figure}  

\clearpage

\begin{figure}[h]
\epsfig{file=2887f2b1.eps, angle=270, width=14cm}
\end{figure}  

\begin{figure}[h]
\epsfig{file=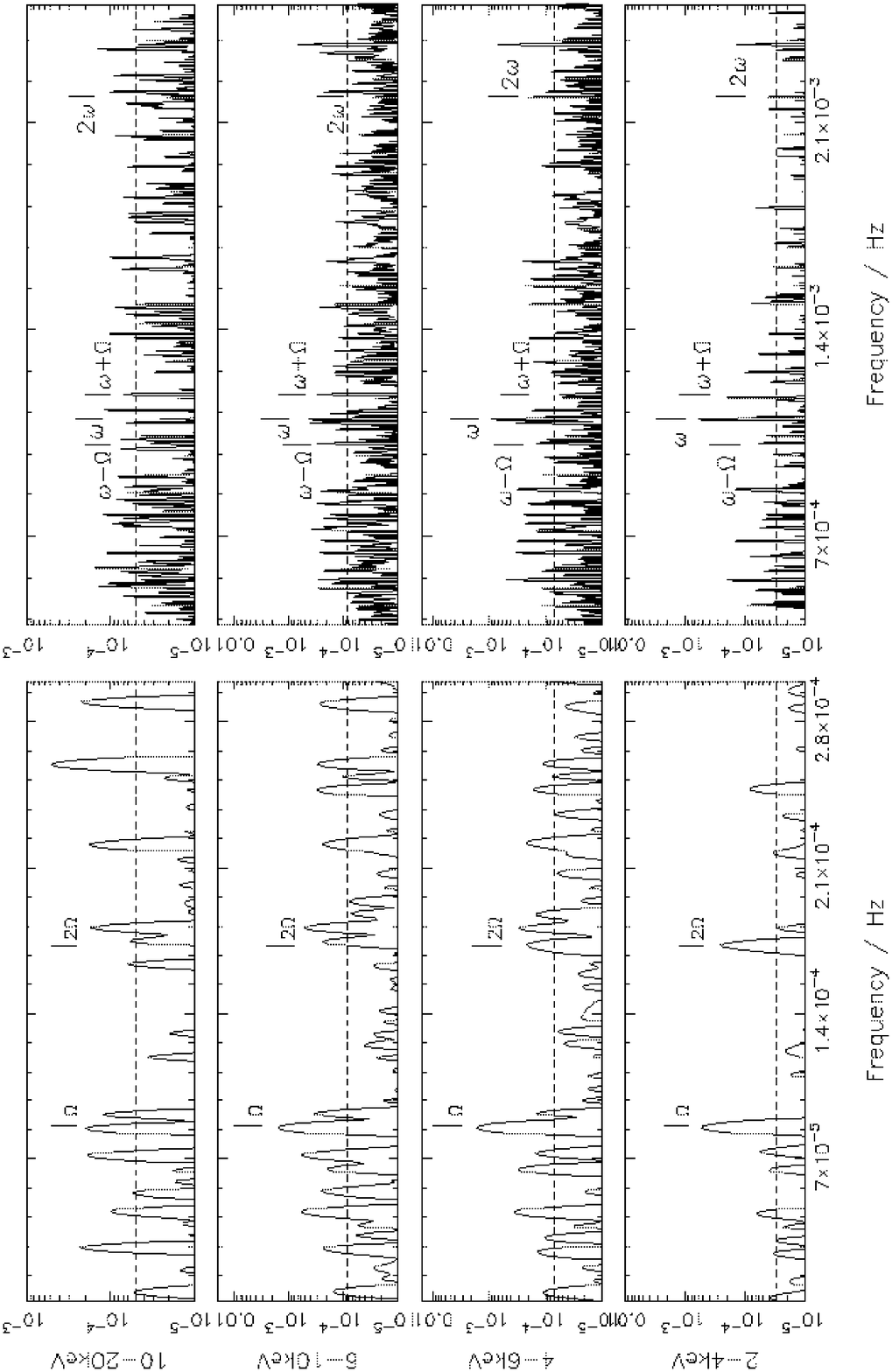, angle=270, width=14cm} 
\setcounter{figure}{1}
\caption{(b) BG CMi}
\end{figure}  

\clearpage

\begin{figure}[h]
\epsfig{file=2887f2c1.eps, angle=270, width=14cm}
\end{figure}  

\begin{figure}[h]
\epsfig{file=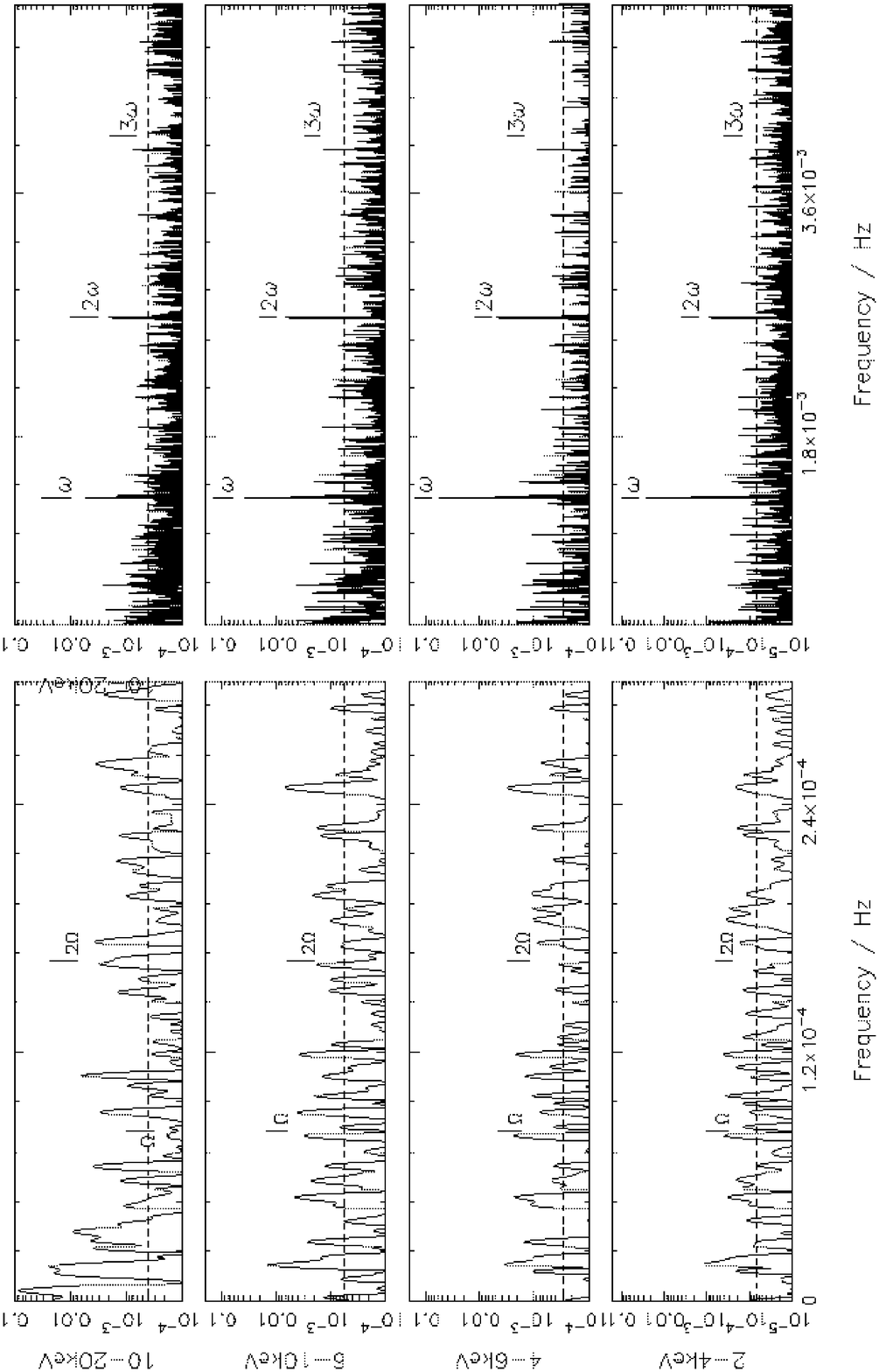, angle=270, width=14cm}
\setcounter{figure}{1}
\caption{(c) V1223 Sgr} 
\end{figure}  

\clearpage

\begin{figure}[h]
\epsfig{file=2887f2d1.eps, angle=270, width=14cm}
\end{figure}  

\begin{figure}[h]
\epsfig{file=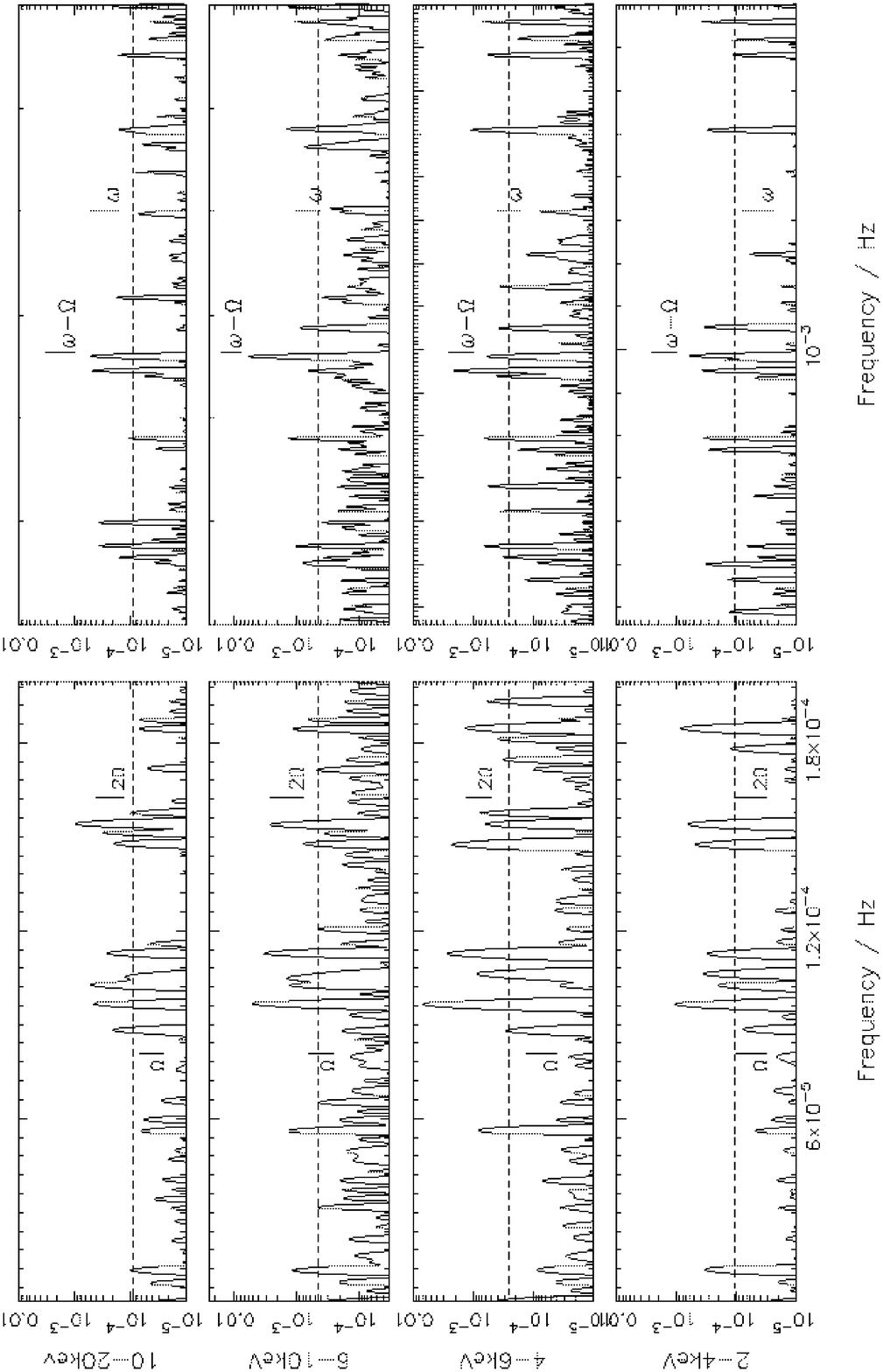, angle=270, width=14cm} 
\setcounter{figure}{1}
\caption{(d) V2400 Oph}
\end{figure}  

\clearpage

\begin{figure}[h]
\epsfig{file=2887f2e1.eps, angle=270, width=14cm}
\end{figure}  

\begin{figure}[h]
\epsfig{file=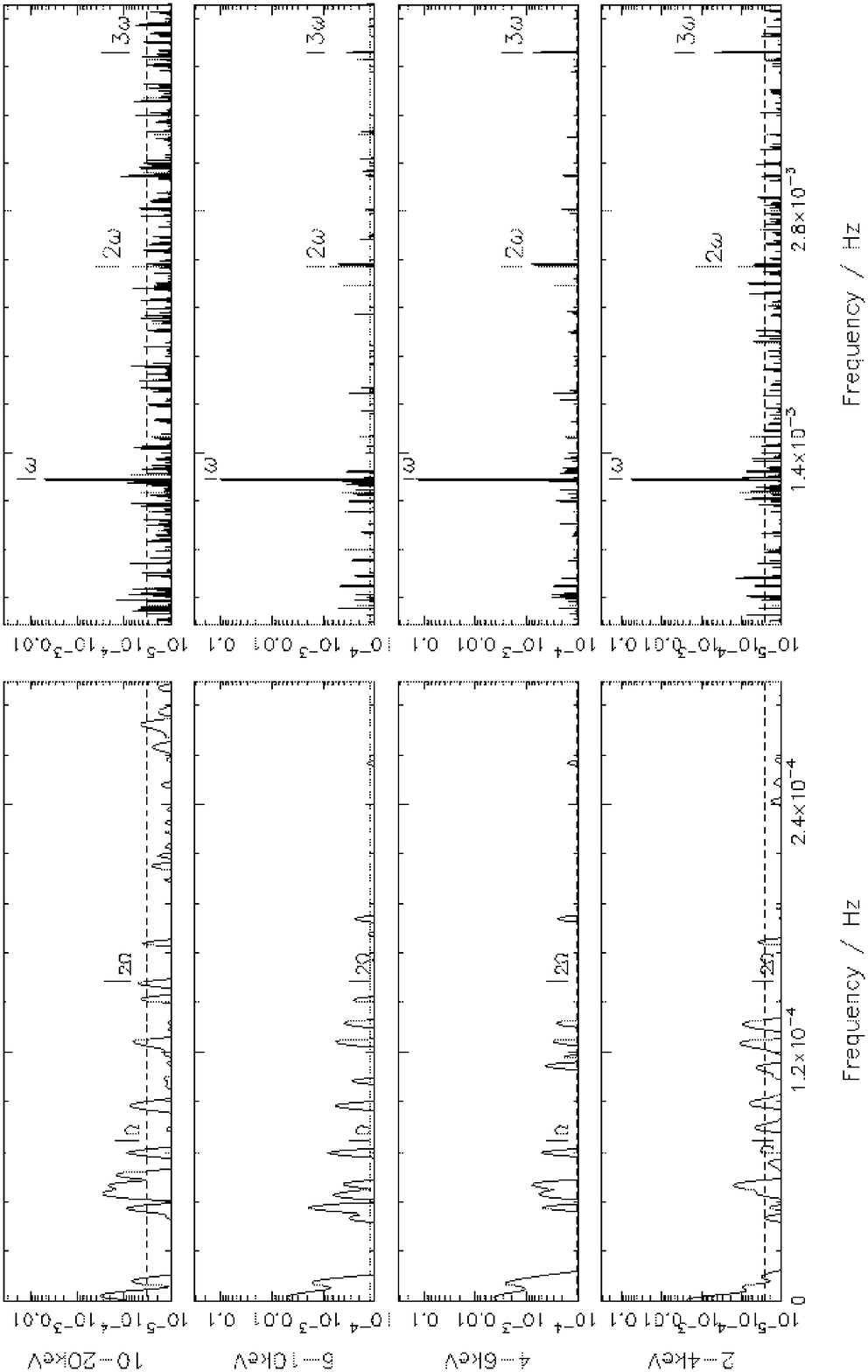, angle=270, width=14cm}
\setcounter{figure}{1}
\caption{(e) AO Psc} 
\end{figure}  

\clearpage

\begin{figure}[h]
\epsfig{file=2887f2f1.eps, angle=270, width=14cm}
\end{figure}  

\begin{figure}[h]
\epsfig{file=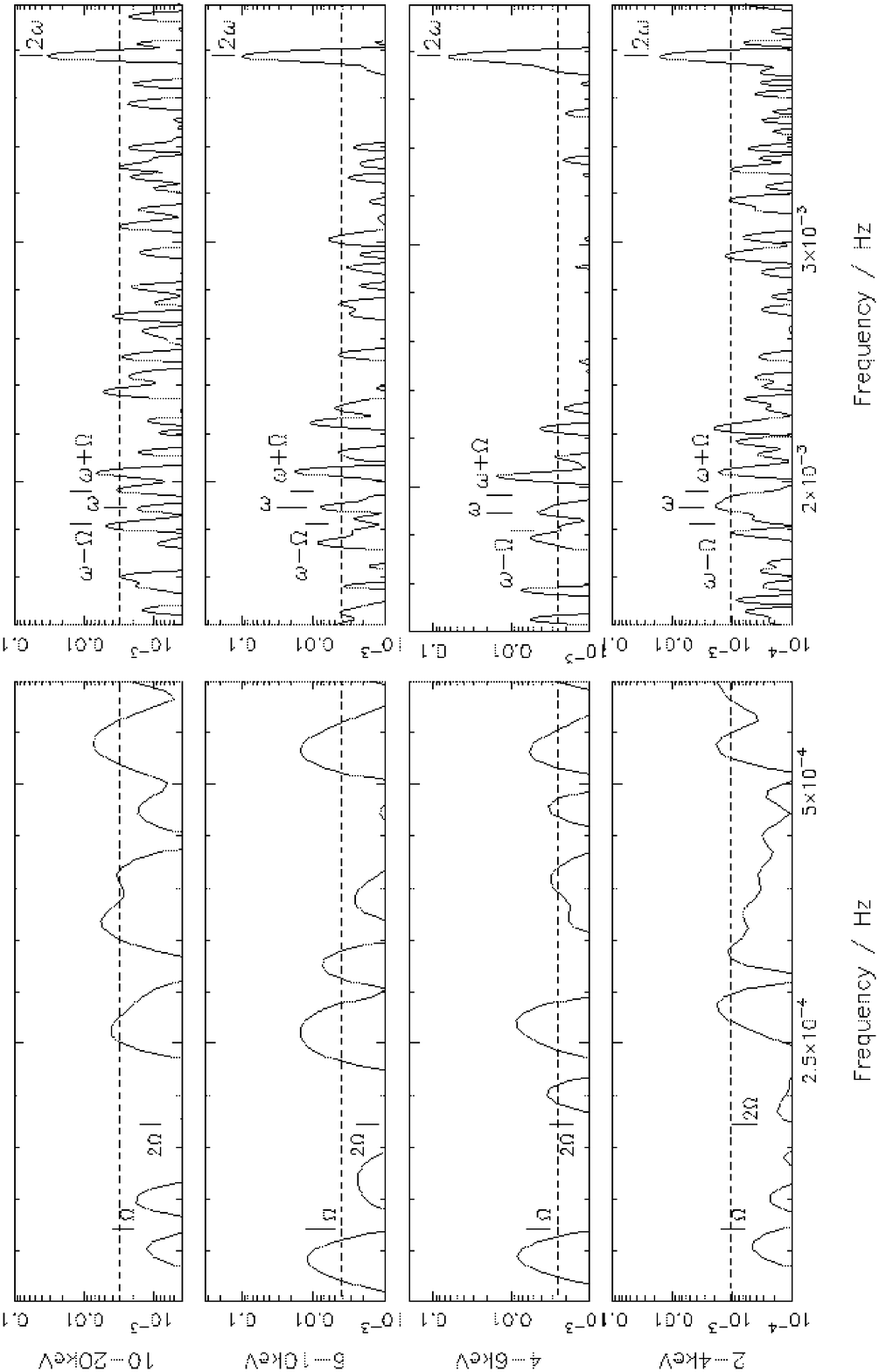, angle=270, width=14cm} 
\setcounter{figure}{1}
\caption{(f) YY Dra}
\end{figure}  

\clearpage

\begin{figure}[h]
\epsfig{file=2887f2g1.eps, angle=270, width=14cm}
\end{figure}  

\begin{figure}[h]
\epsfig{file=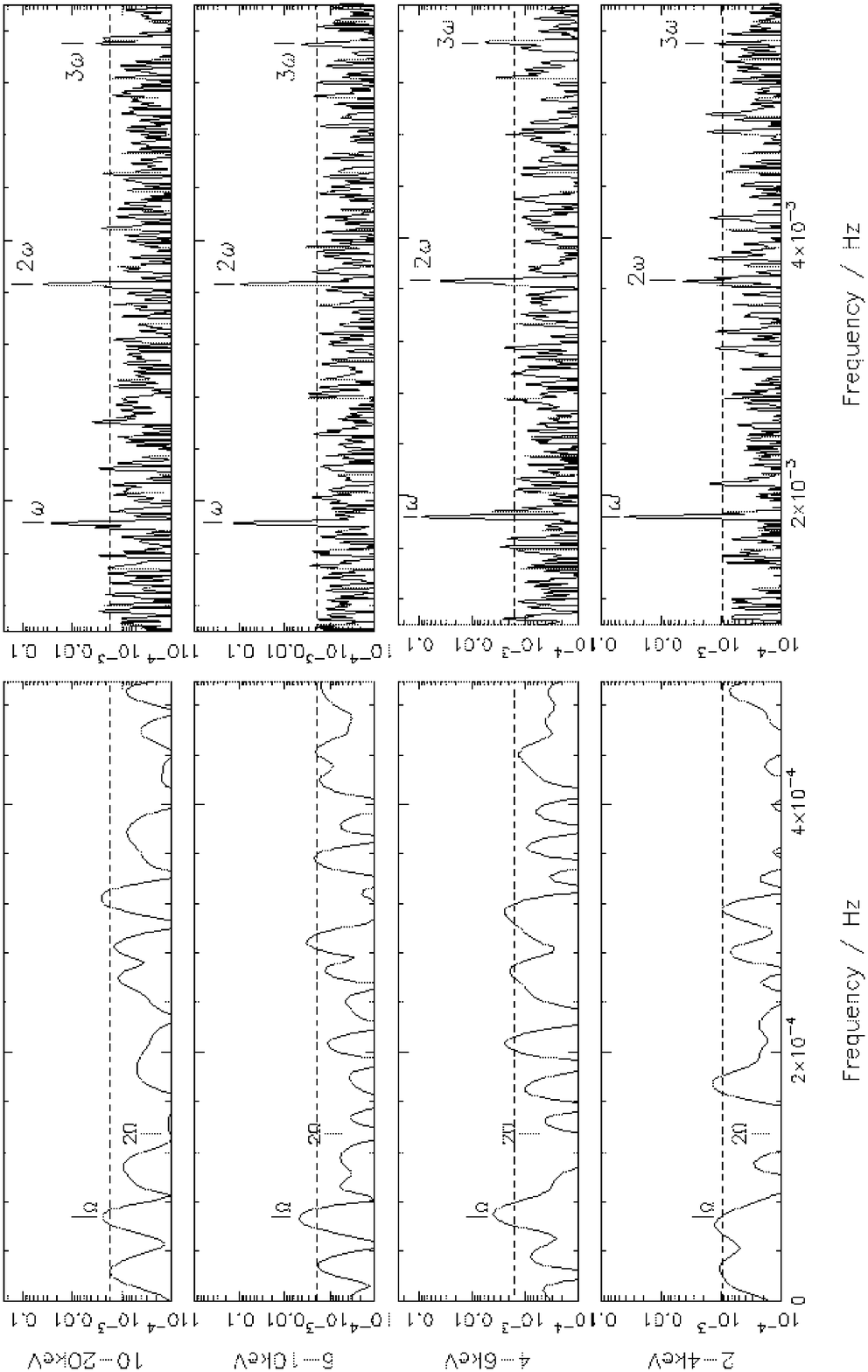, angle=270, width=14cm} 
\setcounter{figure}{1}
\caption{(g) V405 Aur}
\end{figure}  

\clearpage

\begin{figure}[h]
\epsfig{file=2887f2h1.eps, angle=270, width=14cm}
\end{figure}  

\begin{figure}[h]
\epsfig{file=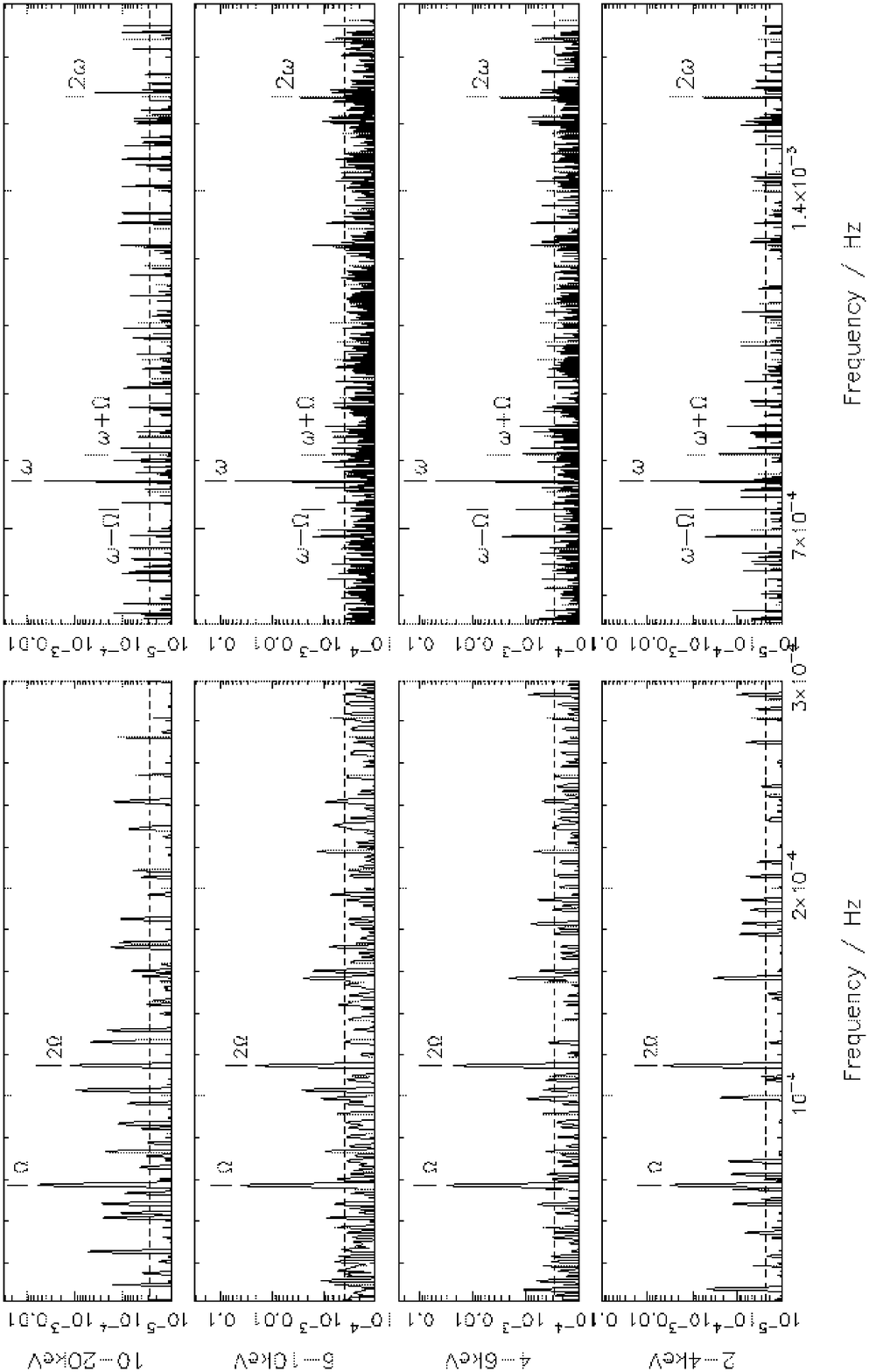, angle=270, width=14cm} 
\setcounter{figure}{1}
\caption{(h) FO Aqr}
\end{figure}  

\clearpage

\begin{figure}[h]
\epsfig{file=2887f2i1.eps, angle=270, width=14cm}
\end{figure}  

\begin{figure}[h]
\epsfig{file=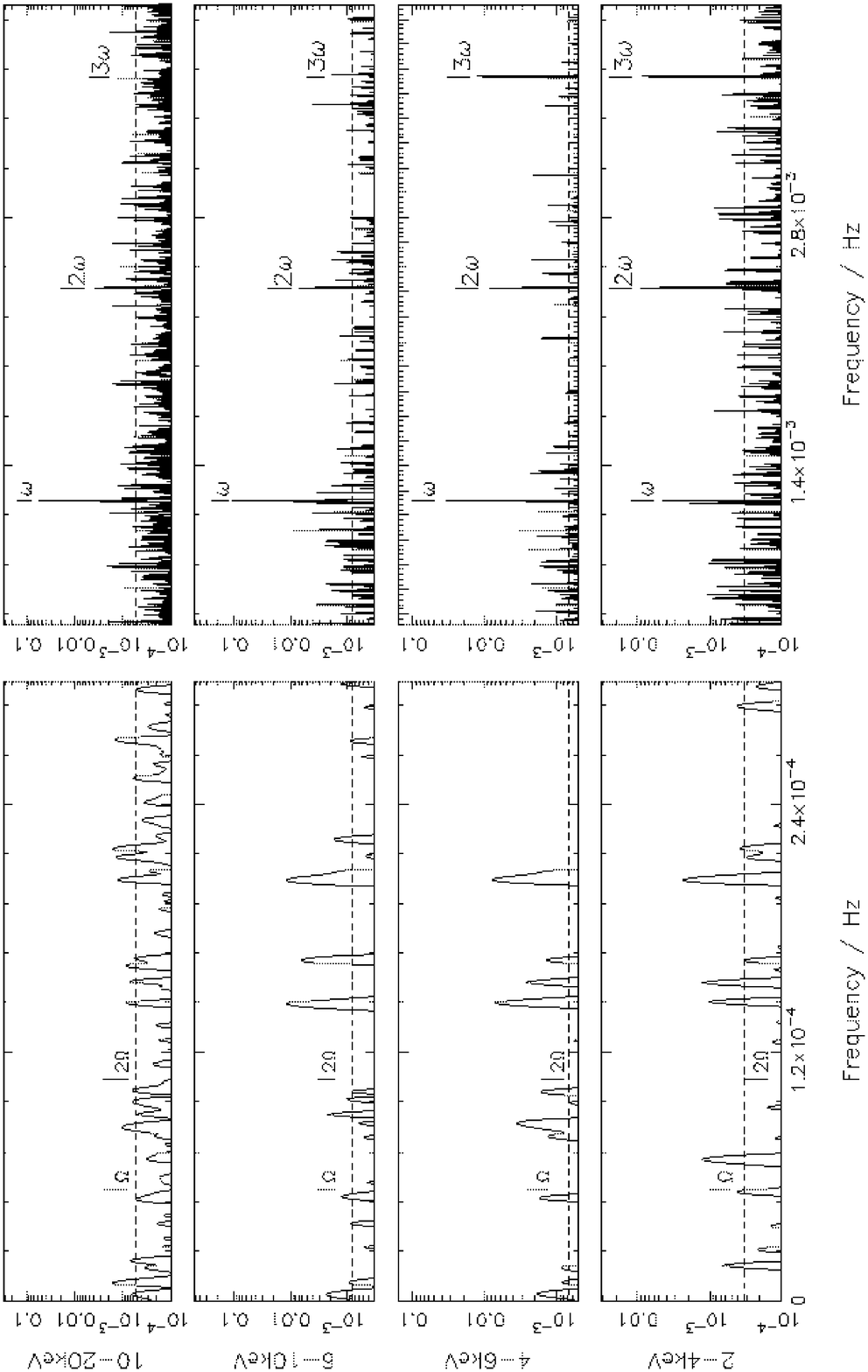, angle=270, width=14cm} 
\setcounter{figure}{1}
\caption{(i) PQ Gem}
\end{figure}  

\clearpage

\begin{figure}[h]
\epsfig{file=2887f2j1.eps, angle=270, width=14cm}
\end{figure}  

\begin{figure}[h]
\epsfig{file=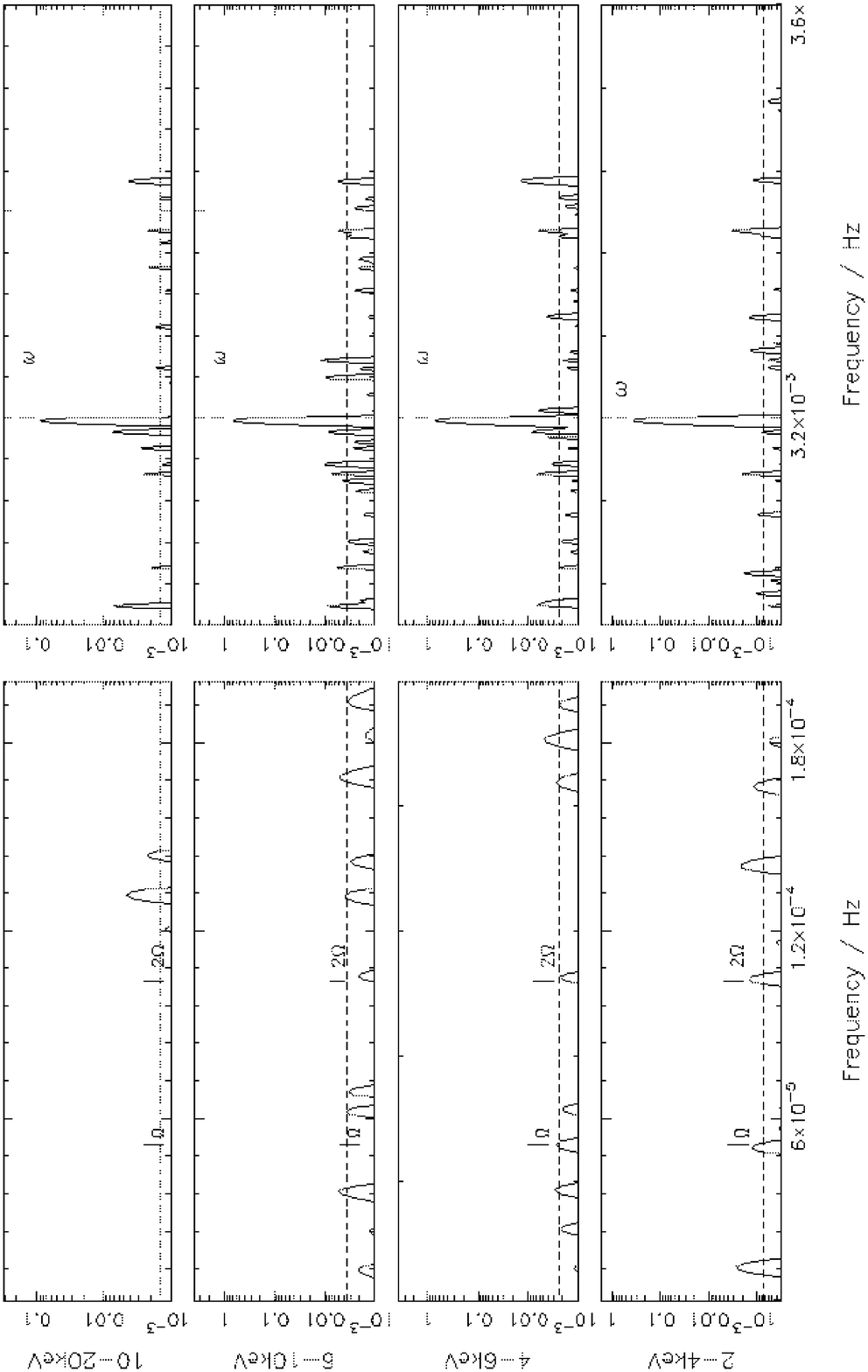, angle=270, width=14cm} 
\setcounter{figure}{1}
\caption{(j) V709 Cas}
\end{figure}  

\clearpage

\begin{figure}[h]
\epsfig{file=2887f2k1.eps, angle=270, width=14cm}
\end{figure}  

\begin{figure}[h]
\epsfig{file=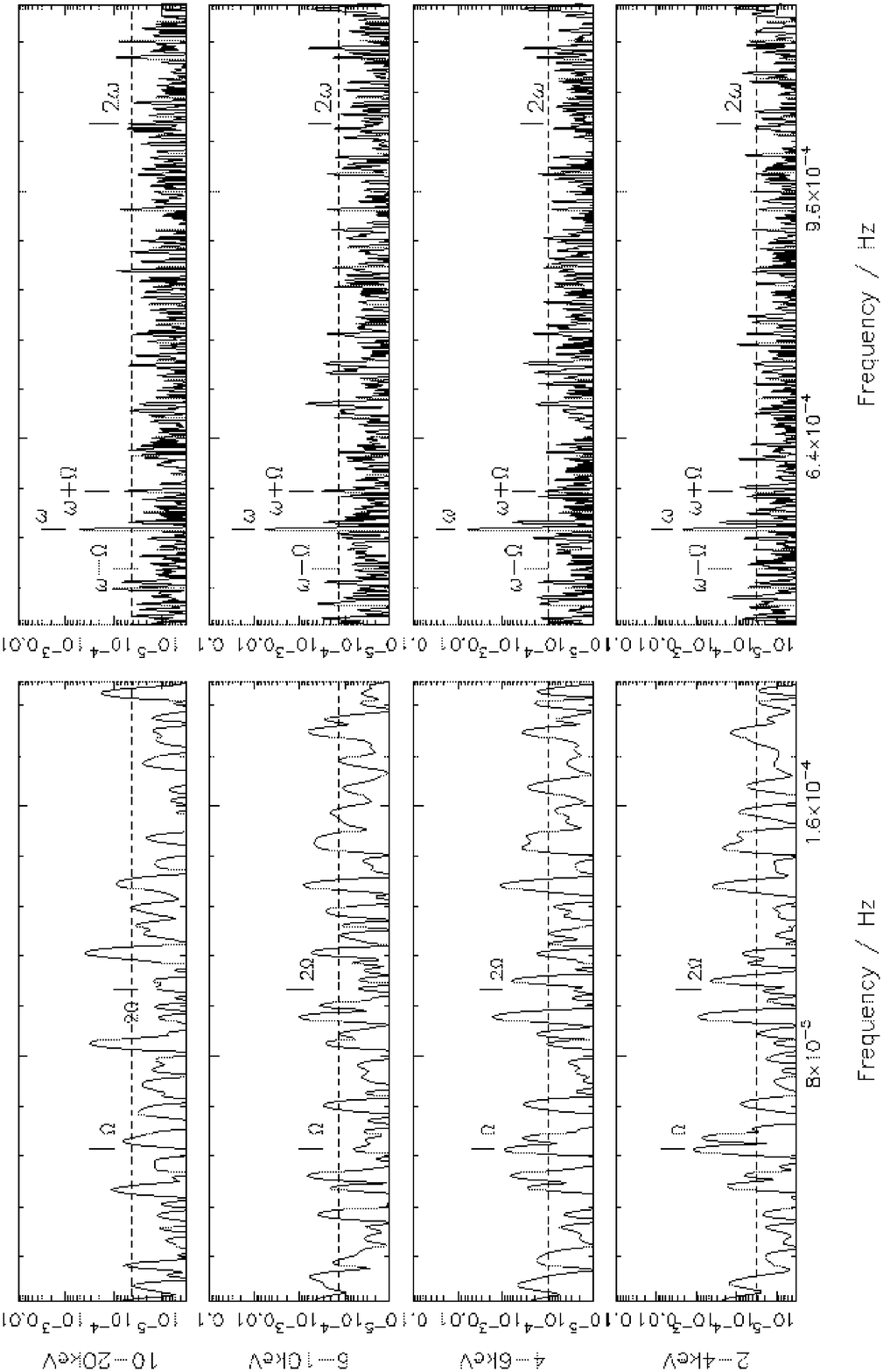, angle=270, width=14cm}
\setcounter{figure}{1}
\caption{(k) TV Col} 
\end{figure}  

\clearpage

\begin{figure}[h]
\epsfig{file=2887f2l1.eps, angle=270, width=14cm}
\end{figure}  

\begin{figure}[h]
\epsfig{file=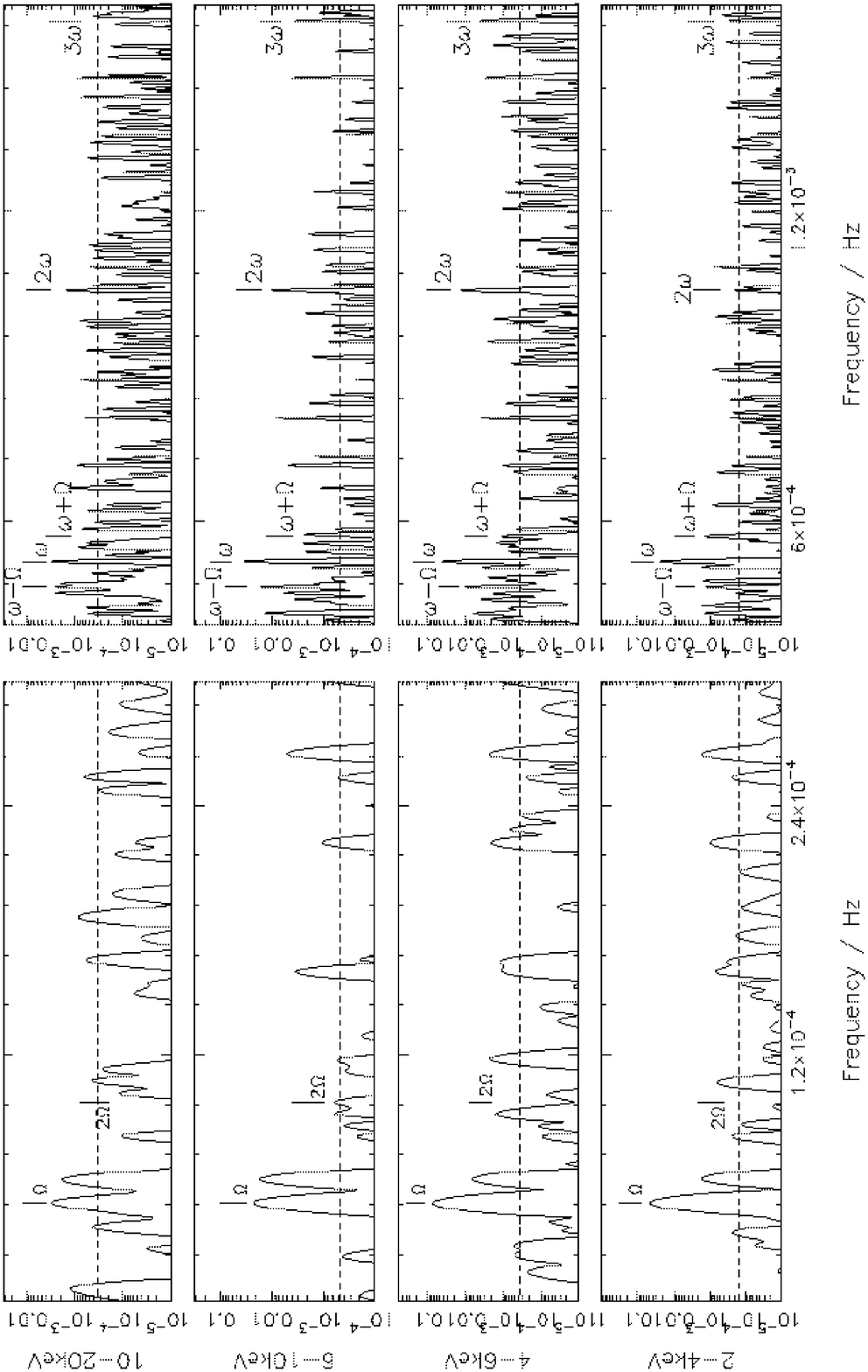, angle=270, width=14cm} 
\setcounter{figure}{1}
\caption{(l) TX Col}
\end{figure}  

\clearpage

\begin{figure}[h]
\epsfig{file=2887f2m1.eps, angle=270, width=14cm}
\end{figure}  

\begin{figure}[h]
\epsfig{file=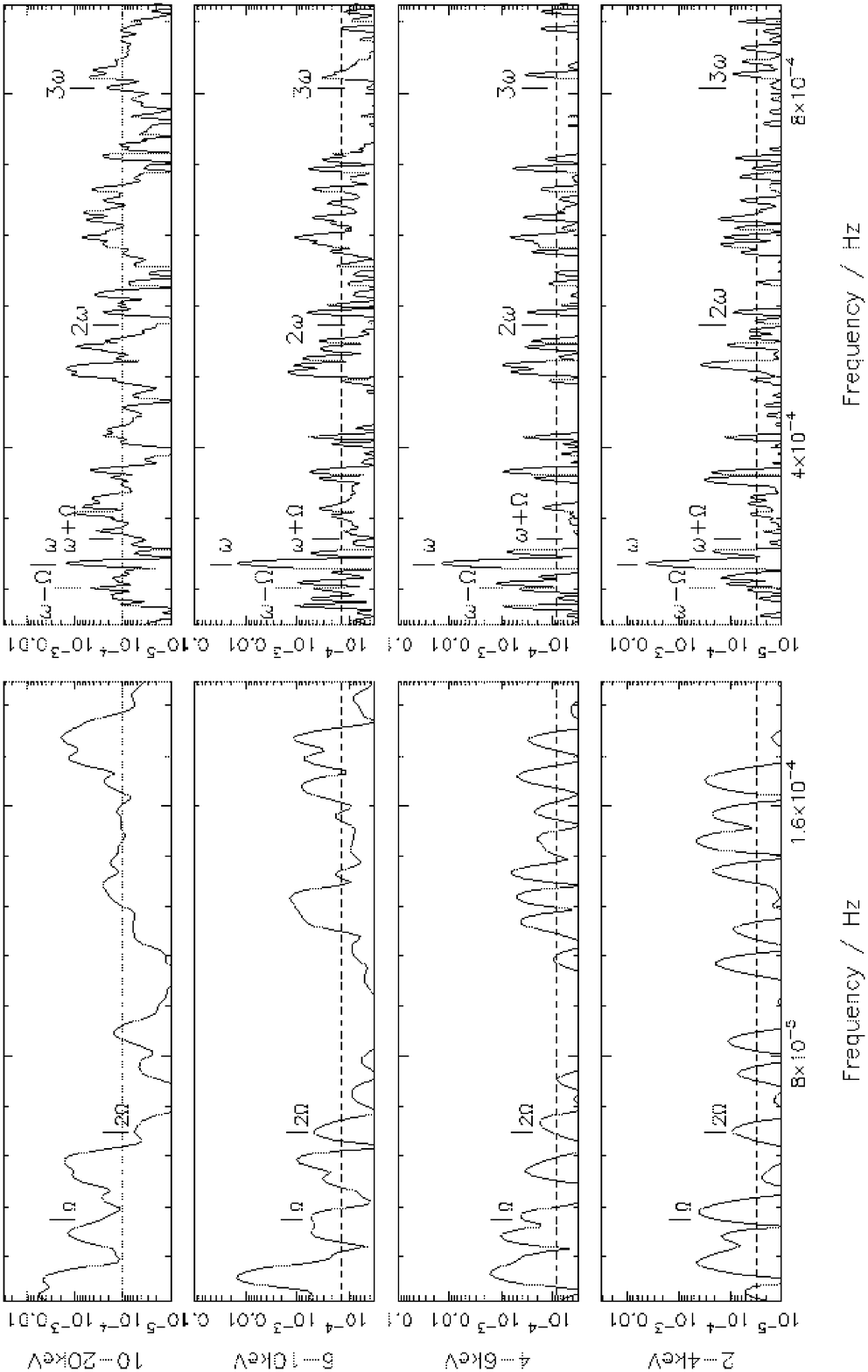, angle=270, width=14cm}
\setcounter{figure}{1}
\caption{(m) V1062 Tau}
\end{figure}  

\clearpage

\begin{figure}[h]
\epsfig{file=2887f3a.eps, angle=270, width=13cm} 
\end{figure}  

\begin{figure}[h]
\epsfig{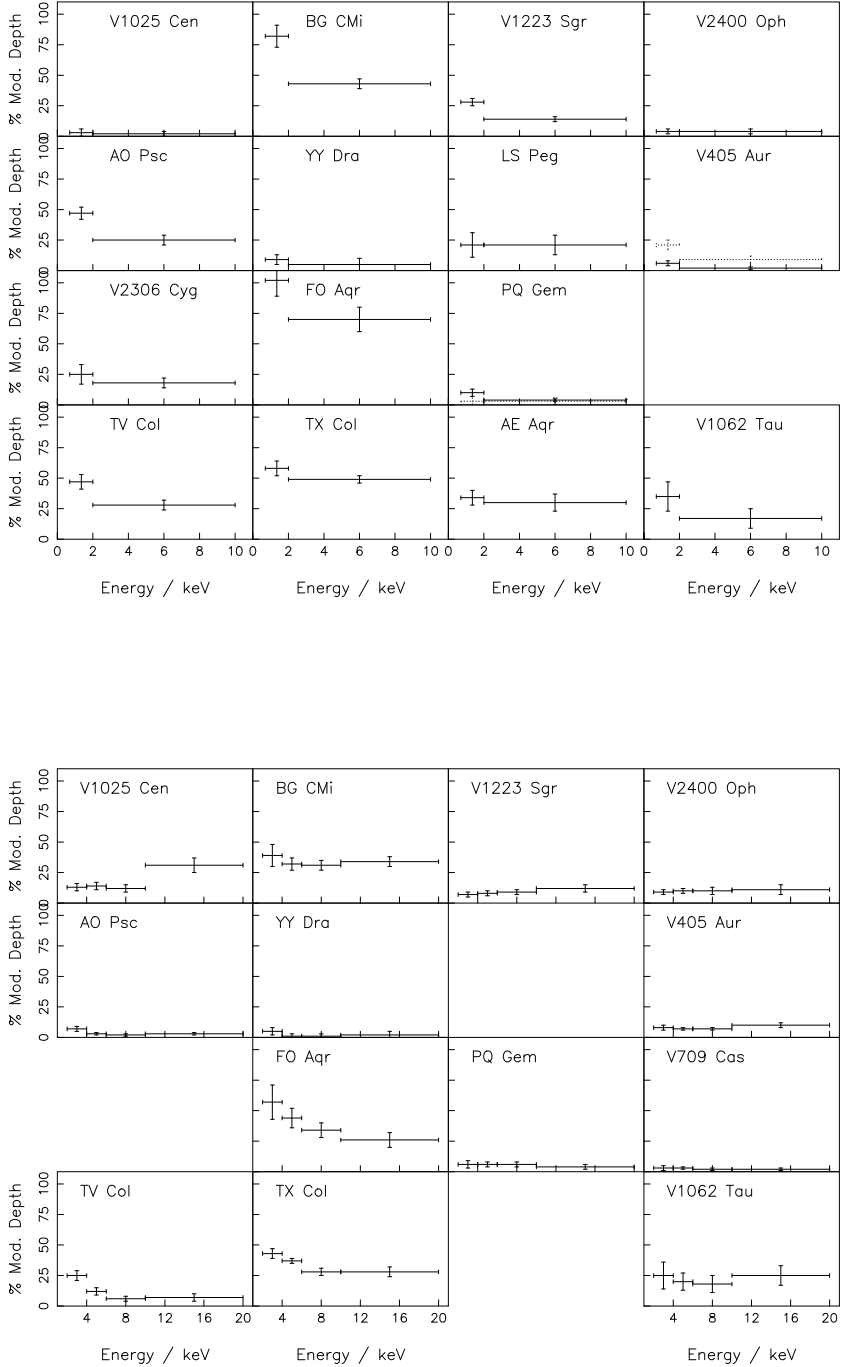} 
\caption{Orbital modulation depths measured from (a) the {\em ASCA} data and 
(b) the {\em RXTE} data. Vertical error bars represent the 1$\sigma$ 
uncertainties in modulation depths, horizontal error bars indicate
the energy range for each measurement. Where more than one observation 
of a source exists, the second observation is illustrated using dotted lines.}
\end{figure}  

\end{document}